\documentclass[aps,prb,reprint]{revtex4-1}

\usepackage{amsmath}
\usepackage{amssymb}
\usepackage{graphicx}
\usepackage{dcolumn}
\usepackage{cases}
\usepackage{mathtools}
\usepackage{dcolumn}
\usepackage{color}
\usepackage{bbold}
\usepackage{bm}
\usepackage{mathrsfs}

\usepackage{lipsum}

\DeclarePairedDelimiter\bra{\langle}{\rvert}
\DeclarePairedDelimiter\ket{\lvert}{\rangle}
\DeclarePairedDelimiter\braket{\langle}{\rangle}

\newcommand\zero{\mathbb{0}}
\newcommand\one{\mathbb{1}}

\renewcommand\vec[1]{\bm{#1}}

\newcommand\RED[1]{{#1}}

\begin{document}

\title{Simple model for electrical hole spin manipulation in semiconductor quantum dots: Impact of dot material and orientation}
\author{Benjamin Venitucci} 
\author{Yann-Michel Niquet}
\email{yniquet@cea.fr}
\affiliation{Université Grenoble Alpes, CEA, IRIG, MEM-L\_Sim, F-38000 Grenoble, France}

\begin{abstract}
We analyze a prototypical particle-in-a-box model for a hole spin qubit. This \RED{quantum dot} is subjected to static magnetic and electric fields, and to a radio-frequency electric field that drives Rabi oscillations owing to spin-orbit coupling. We derive the equations for the Rabi frequency in a regime where the Rabi oscillations mostly result from the coupling between the qubit states and a single nearby \RED{excited state}. This regime has been shown to prevail in, e.g., hole spin qubits in thin silicon-on-insulator nanowires. The equations for the Rabi frequency highlight the parameters that control the Rabi oscillations. We show, in particular, that $[110]$-oriented \RED{dots} on $(001)$ substrates perform much better than $[001]$-oriented \RED{dots} because they take best advantage of the anisotropy of the valence band of the host material. We also conclude that silicon provides the best opportunities for fast Rabi oscillations in this regime despite small spin-orbit coupling.
\end{abstract}

\maketitle

Spins in semiconductor quantum dots are an attractive platform for quantum information technologies.\cite{Kane98,Loss98} Electron spin quantum bits (qubits) have, in particular, been demonstrated in different III-V materials over the last two decades,\cite{Petta05,Koppens06,Hanson07} and much more recently in silicon.\cite{Pla12} Silicon\cite{Zwanenburg13} is indeed a promising host material for spin qubits as it can be isotopically purified from the nuclear spins that may interact with the electron spins. Very long spin coherence times,\cite{Tyryshkin12} as well as single and two qubit gates with high fidelity have thereby been reported in silicon. The quantum dots in these devices are defined by an impurity, by electrostatics and/or by lithography.\cite{Veldhorst14,Veldhorst15b,Takeda16,Yoneda18,Watson18}

Hole spin qubits have also been proposed and successfully demonstrated in the last few years.\cite{Kloeffel13,Maurand16,Crippa18,Katsaros18,Crippa19} Hole spins are much more efficiently coupled to the orbital motion of the carrier than electron spins. This spin-orbit coupling (SOC) is a relativistic effect that can be described semi-classically as the action of the magnetic field created by the nuclei moving in the frame of a carrier onto its spin.\cite{Winkler03} It is stronger for holes than for electrons because the Bloch functions of the top of the valence band are essentially degenerate combinations of atomic $p$-orbitals, which are tightly coupled to the spin by the intra-atomic SOC Hamiltonian $H_{\rm SOC}\propto\vec{L}\cdot\vec{S}$ ($\vec{L}$ and $\vec{S}$ being respectively the atomic angular momentum and spin operators). Strong SOC might enhance the interactions of the spins with electrical noise and phonons, hence speed-up decoherence; however it provides outstanding opportunities for very fast, all-electrical manipulation by Electric Dipole Spin Resonance (EDSR).\cite{Rashba03,Kato03,Golovach06,Flindt06,Nowack07,Rashba08,NadjPerge10,vandenBerg13,Maurand16,Corna18,Crippa18,Katsaros18}

EDSR on hole spins has, for example, been demonstrated in silicon-on-insulator (SOI) devices.\cite{Maurand16,Crippa18,Crippa19} The quantum dot is there defined electrostatically by a gate lying on top of an etched nanowire. A radio-frequency modulation of the voltage on that gate drives Rabi oscillations of the hole spin with frequencies as large as a few tens to a hundred of MHz. The hole spins show rich physics, as highlighted by the complex dependence of the Rabi frequency on the orientation of the static magnetic field.\cite{Crippa18} This dependence was shown to result from a complex interplay between the effects of the motion of the dot as a whole in the electric field of the gate and the changes in the shape of that dot brought by the anharmonic components of the potential. These mechanisms can be described by a unified framework based on the measurement or calculation of a gyromagnetic $g$-matrix and of its derivative with respect to the gate voltage.\cite{Crippa18,Venitucci18}

We have used this $g$-matrix formalism to simulate realistic SOI hole devices and rationalize the dependence of the Rabi frequency on the orientation of the magnetic field.\cite{Venitucci18} We have, in particular, shown that the Rabi oscillations essentially result from the coupling of the qubit states (with a mostly ``$s$-like'' envelope) with a nearby \RED{excited state} (with a mostly ``$p$-like'' envelope) under a combination of electric and magnetic fields that breaks time-reversal symmetry. In the present work, we propose a prototypical model for this regime, based on a box subjected to homogeneous electric and magnetic fields. The model can be solved analytically and the equations highlight the mechanisms and parameters that control the Rabi oscillations. In particular, we show that thin $[110]$-oriented box on $(001)$ substrates perform much better than thin $[001]$-oriented box because they take best advantage of the anisotropy of the valence band of the host material. Also, we conclude that silicon hole qubits are expected to exhibit the fastest Rabi oscillations in this regime as this material displays the most anisotropic valence band (among conventional semiconductors), despite smaller spin-orbit coupling.

We introduce the model in section \ref{sectionModel}, then compute the Rabi frequency of the hole qubit in section \ref{sectionRabi}; Finally, we discuss the physics and the dependence of the Rabi frequency on the material and quantum dot parameters in section \ref{sectionDiscussion}.

\section{Model}
\label{sectionModel}

In this section, we introduce the model for the box, then the Luttinger-Kohn, four bands $\vec{k}\cdot\vec{p}$ Hamiltonian used to describe the electronic structure of the holes. We next discuss the solution of this Hamiltonian in a minimal basis set capturing the main physics. We analyze, in particular, the effects of quantum confinement, electric and magnetic fields, in order to prepare the calculation of the Rabi frequency of the hole qubit in section \ref{sectionRabi}.

\subsection{System}
\label{section_system}
\begin{figure}
\includegraphics[width=0.95\columnwidth]{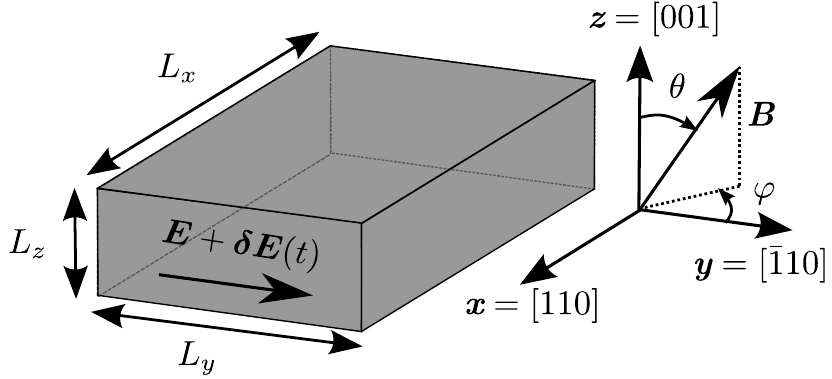}
\caption{The model system. A \RED{rectangular} box with sides $L_x$, $L_y$ and $L_z$ is subjected to a static magnetic field $\vec{B}$, a static electric field $\vec{E}=E_0\vec{y}$ and a radio-frequency electric field modulation $\delta\vec{E}(t)=E_{\rm ac}\sin(2\pi f_Lt+\phi)\vec{y}$. The orientation of $\vec{B}$ is characterized by the polar angle $\theta$ and the azimuthal angle $\varphi$.}
\label{figqubit}
\end{figure}

We consider a \RED{rectangular} box with sides $L_x$, $L_y$ and $L_z$ along axes $x\parallel[110]$, $y\parallel[\bar{1}10]$, and $z\parallel[001]$ (other orientations will be discussed in section \ref{sectionDiscussion}). We assume a hard wall confinement potential:
\begin{equation}
V_{\rm box}(x,y,z)=
\begin{cases} 
0\text{ if }|x|<\frac{L_x}{2},\,|y|<\frac{L_y}{2},\,|z|<\frac{L_z}{2}\,,\\
+\infty \text{ otherwise.}
\end{cases}
\end{equation}
The box is subjected to a static magnetic field $\vec{B}$ and to a static electric field $\vec{E}=E_0\vec{y}$ applied by external gates (see Fig. \ref{figqubit}). The same gates will be used to drive Rabi oscillations in section \ref{sectionRabi}. In addition, the box may undergo in-plane biaxial strain $\varepsilon_{xx}=\varepsilon_{yy}=\varepsilon_\parallel$, $\varepsilon_{zz}=\varepsilon_\perp=-\nu\varepsilon_\parallel$, where $\nu=2c_{12}/c_{11}$ is the biaxial Poisson ratio and $c_{11}$, $c_{12}$ are the elastic constants of the box material. 

This model is meant to be the simplest description of a hole spin qubit as implemented in planar and SOI devices. As shown below, it captures the main physics outlined in the simulations of Ref. \onlinecite{Venitucci18}.

\subsection{Luttinger-Kohn Hamiltonian}

We assume that the holes in the box can be described by the four bands Luttinger-Kohn (LK) Hamiltonian.\cite{Luttinger55,KP09} In the bulk material, the degenerate heavy- and light-hole Bloch functions at $\Gamma$ can be mapped onto the eigenstates $\ket{j_z}$ of an angular momentum $J=3/2$. The LK Hamiltonian then reads in the $\{\ket{+\frac{3}{2}},\ket{+\frac{1}{2}},\ket{-\frac{1}{2}},\ket{-\frac{3}{2}}\big\}$ basis set:
\begin{equation}
H_{\rm LK}=
\begin{pmatrix}
P+Q & -S & R & 0 \\
-S^* & P-Q & 0 & R \\
R^* & 0 & P-Q & S \\
0 & R^* & S^* & P+Q \\
\end{pmatrix}\,,
\label{hamiltonian_without_field}
\end{equation}
where:
\begin{subequations}
\begin{align}
P&=\frac{\hbar^2}{2m_0}\gamma_1\Big(k_x^2+k_y^2+k_z^2\Big) \\
Q&=\frac{\hbar^2}{2m_0}\gamma_2\Big(k_x^2+k_y^2-2k_z^2\Big) \\
R&=\frac{\hbar^2}{2m_0}\sqrt{3}\Big[-\gamma_3\Big(k_x^2-k_y^2\Big)+2i\gamma_2k_xk_y\Big] \label{eqR} \\
S&=\frac{\hbar^2}{2m_0}2\sqrt{3}\gamma_3\Big(k_x-ik_y\Big)k_z\,.
\end{align}
\end{subequations}
$\vec{k}=(k_x, k_y, k_z)$ is the wave vector, $m_0$ is the free electron mass, and $\gamma_1$, $\gamma_2$, $\gamma_3$ are the Luttinger parameters that characterize the anisotropic mass of the holes. Strains are dealt with in Appendix \ref{AppendixStrains}. Note that we assume positive (electron-like) dispersion for the holes for the sake of simplicity, and that we discard the ``indirect'' Dresselhaus and Rashba spin-orbit interactions that may arise from the coupling with remote Bloch functions owing to the breaking of inversion symmetry by the lattice (III-V materials), by the static electric field and by the interfaces.\cite{Winkler03} The Rabi oscillations of hole spins are indeed expected to be dominated by the ``direct'' spin-orbit interaction within the heavy- and light-holes manifold.\cite{Kloeffel11,Kloeffel13,Kloeffel18}

\subsection{(Minimal) basis set for the envelope functions}
\label{minimal_basis}

In the box (assuming at first zero electric and magnetic field), the substitution $\vec{k}\rightarrow-i\vec{\nabla}$ in the LK Hamiltonian yields a set of four coupled differential equations for the heavy-hole ($j_z=\pm 3/2$) and light-hole ($j_z=\pm 1/2$) envelope functions. We expand the eigensolutions of these equations in the basis of harmonic functions $\{\ket{n_xn_yn_z}\otimes\ket{j_z}\}$, where $\braket{\vec{r}|n_xn_yn_z}=\chi_{n_x}(x, L_x)\chi_{n_y}(y, L_y)\chi_{n_z}(z, L_z)$ and:
\begin{equation}
\chi_n(u, L) = \sqrt{\frac{2}{L}}\sin\Big[n \pi\Big(\frac{u}{L}+\frac{1}{2}\Big)\Big]\,,|u|\le\frac{L}{2}\,.
\end{equation}
The Hamiltonian can be diagonalized numerically in the above basis set; However, in order to highlight trends in material and device parameters, it is instructive to build a ``minimal'' model that captures the essential physics. \RED{The ground-state of the Hamiltonian turns out to be mostly ``$s$-like'' ($n_x=n_y=n_z=1$). When $L_z\ll L_x,\,L_y$ (a ``thin dot'' limit we will focus on later), the lowest-lying excited states involve envelopes with increasing quantum numbers $n_x$ and $n_y$. However, envelopes with $n_x>1$ play little role in the present model as the electric field in the box is oriented along $\vec{y}$. As a matter of fact, the $s$-like ground-state gets mostly mixed with a ``$p_y$-like'' excitation ($n_y=2$) by this electric field.} Therefore, we will establish analytical results in the following minimal basis set that includes the heavy- and light-hole $s$ and $p_y$ envelopes:
\begin{equation}
\mathcal{B}=\{\mathcal{B}_0,T\mathcal{B}_0\}
\end{equation}
where:
\begin{equation}
\mathcal{B}_0=\Big\{\ket{1,+\frac{3}{2}},\ket{1,-\frac{1}{2}},\ket{2,+\frac{3}{2}},\ket{2,-\frac{1}{2}}\Big\}\,,
\end{equation}
$\ket{i,j_z}=\ket{1i1}\otimes\ket{j_z}$, and $T$ is the time-reversal symmetry operator ($T\ket{+\frac{3}{2}}=\ket{-\frac{3}{2}}$ and $T\ket{-\frac{1}{2}}=\ket{+\frac{1}{2}}$). We discuss the effects of structural confinement, electric and magnetic fields in this basis set in the next paragraphs.

\subsection{Effects of structural confinement}

\begin{figure}
\includegraphics[width=0.95\columnwidth]{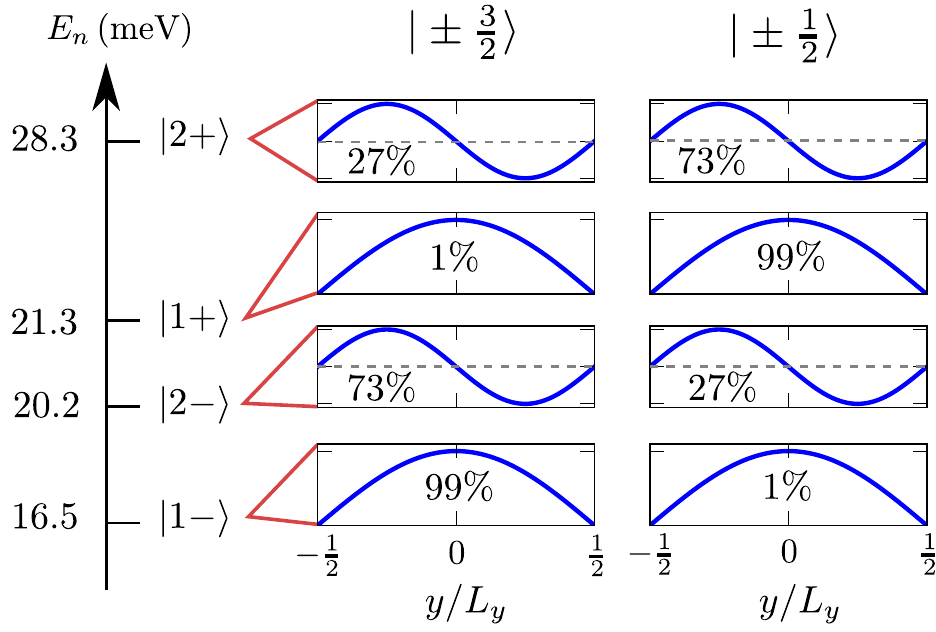}
\caption{Energy levels and envelope functions of a silicon dot with sides $L_x=40$ nm, $L_y=30$ nm and $L_z=10$ nm (at zero electric and magnetic fields). The total weight of each envelope is indicated in the corresponding panel. The Luttinger parameters of silicon are given in Table \ref{tabmat}.}
\label{figdotSi}
\end{figure}

In the basis set $\mathcal{B}$, neither $S$ nor the $\propto\gamma_2$ component of $R$ do contribute to the matrix elements of $H_{\rm LK}$ as all basis functions have the same \RED{quantum numbers $n_x$ and $n_z$}. Therefore, the Hamiltonian is block diagonal at zero fields:
\begin{equation}
\mathcal{H}_{\rm LK}=
\begin{pmatrix}
\mathcal{H}_0 & 0_{4\times4}\\
0_{4\times4} & \mathcal{H}_0
\end{pmatrix}\,,
\label{eqHLK}
\end{equation}
where in the basis set $\mathcal{B}_0$:
\begin{equation}
\mathcal{H}_0= 
\begin{pmatrix}
P_1+Q_1 & R_1 & 0 & 0 \\
R_1 & P_1-Q_1 & 0 & 0 \\
0 & 0 & P_2+Q_2 & R_2 \\
0 & 0 & R_2 & P_2-Q_2 \\
\end{pmatrix}\,,
\end{equation}
with:
\begin{subequations}
\begin{align}
P_1&= \frac{\hbar^2}{2m_0}\gamma_1\pi^2\Big(L_x^{-2}+L_y^{-2}+L_z^{-2}\Big) \\
Q_1&= \frac{\hbar^2}{2m_0}\gamma_2\pi^2\Big(L_x^{-2}+L_y^{-2}-2L_z^{-2}\Big) \\
R_1&=-\frac{\hbar^2}{2m_0}\sqrt{3}\gamma_3\pi^2\Big(L_x^{-2}-L_y^{-2}\Big)\,,
\end{align}
\end{subequations}
and:
\begin{subequations}
\begin{align}
P_2&= \frac{\hbar^2}{2m_0}\gamma_1\pi^2\Big(L_x^{-2}+4L_y^{-2}+L_z^{-2}\Big) \\
Q_2&= \frac{\hbar^2}{2m_0}\gamma_2\pi^2\Big(L_x^{-2}+4L_y^{-2}-2L_z^{-2}\Big) \\
R_2&=-\frac{\hbar^2}{2m_0}\sqrt{3}\gamma_3\pi^2\Big(L_x^{-2}-4L_y^{-2}\Big)\,.
\end{align}
\end{subequations}
The $n_y=1$ heavy-hole state is therefore mixed with the $n_y=1$ light-hole state by $R_1$, while the $n_y=2$ heavy-hole state is mixed with the $n_y=2$ light-hole state by $R_2$. These couplings are driven by lateral confinement ($R_i\propto L_x^{-2},\,L_y^{-2}$). The eigenstates $\ket{i\pm}$ ($i\equiv n_y=1,\,2$) of $\mathcal{H}_0$ are actually:
\begin{subequations}
\label{eqipmnofield}
\begin{align}
\ket{i-}&=h_i\ket{i,+\frac{3}{2}}+l_i\ket{i,-\frac{1}{2}} \\
\ket{i+}&=-l_i\ket{i,+\frac{3}{2}}+h_i\ket{i,-\frac{1}{2}}\,,
\end{align}
\end{subequations}
where $h_i=-R_i/W_i$, $l_i=(Q_i+\sqrt{Q_i^2+R_i^2})/W_i$, and $W_i^2=R_i^2+(Q_i+\sqrt{Q_i^2+R_i^2})^2$. The associated eigenenergies are:
\begin{equation}
E_{i\pm}=P_i\pm\sqrt{Q_i^2+R_i^2}\,.
\end{equation}
The $\ket{i-}$ states are dominated by the heavy-hole $\ket{i,+\frac{3}{2}}$ component in the \RED{thin dot limit $L_z\ll L_x,\,L_y$}, while the $\ket{i+}$ states are dominated by the light-hole $\ket{i,-\frac{1}{2}}$ component. This is illustrated in Fig. \ref{figdotSi}, which shows the energy levels and envelope functions of a silicon dot with sides $L_x=40$ nm, $L_y=30$ nm and $L_z=10$ nm. Note that the $\ket{1\pm}$ remain pure heavy- and light-hole states when $L_y=L_x$ ($R_1=0$), while the $\ket{2\pm}$ states remain so when $L_y=2L_x$ ($R_2=0$). 

Each of the $\ket{i\pm}$ state is twice degenerate owing to time-reversal symmetry [see Eq. (\ref{eqHLK})]. The degenerate partner in the $T\mathcal{B}_0$ basis set has the same expression as Eq. (\ref{eqipmnofield}) with $\ket{i,+\frac{3}{2}}$ replaced with $\ket{i,-\frac{3}{2}}$ and $\ket{i,-\frac{1}{2}}$ replaced with $\ket{i,+\frac{1}{2}}$. We therefore introduce a pseudo-spin index to distinguish the $\ket{i\pm,\Uparrow}$ states in the $\mathcal{B}_0$ basis set [Eqs. (\ref{eqipmnofield})] from their degenerate, time-reversal symmetric counterparts $\ket{i\pm,\Downarrow}$ in the $T\mathcal{B}_0$ basis set.

\subsection{Effects of the static electric field}
\label{electric_field_section}
The Hamiltonian of the potential $V_{\rm e}=-eE_0y$ associated with the static electric field $\vec{E}=E_0\vec{y}$ is diagonal with respect to the angular momentum $j_z$, and takes the following form in the basis sets $\mathcal{B}_0$ and $T\mathcal{B}_0$:
\begin{equation}
\mathcal{H}_{\rm e}=\Lambda
\begin{pmatrix}
0 & 0 & 1 & 0 \\
0 & 0 & 0 & 1 \\
1 & 0 & 0 & 0 \\
0 & 1 & 0 & 0 \\
\end{pmatrix}\,,
\end{equation}
where $\Lambda=16eE_0L_y/(9\pi^2)$. Therefore, as discussed above, the electric field mixes the $n_y=1$ and $n_y=2$ states with the same $j_z$.

In order to achieve analytical results, we shall, in a first approximation, deal with the static electric field to first order in perturbation. We hence introduce the first-order $s$-like states:
\begin{subequations}
\label{eqtilde1}
\begin{align}
\ket{\tilde{1}-}&=\ket{1-}+\lambda_{2-}^{1-}\ket{2-}+\lambda_{2+}^{1-}\ket{2+} \\
\ket{\tilde{1}+}&=\ket{1+}+\lambda_{2-}^{1+}\ket{2-} +\lambda_{2+}^{1+}\ket{2+}\,,
\end{align}
\end{subequations}
and the first-order $p_y$-like states:
\begin{subequations}
\label{eqtilde2}
\begin{align}
\ket{\tilde{2}-}&=\ket{2-}+\lambda_{1-}^{2-}\ket{1-}+\lambda_{1+}^{2-}\ket{1+} \\
\ket{\tilde{2}+}&=\ket{2+}+\lambda_{1-}^{2+}\ket{1-}+\lambda_{1+}^{2+}\ket{1+}\,,
\end{align}
\end{subequations}
where:
\begin{subequations}
\label{eqlambdas}
\begin{align}
\lambda_{2\pm}^{1\pm}&=-\lambda_{{1\pm}}^{{2\pm}}=\hphantom{\pm}\Lambda\frac{h_1h_2+l_1l_2}{E_{{1\pm}}-E_{{2\pm}}} \\ 
\lambda_{2\mp}^{1\pm}&=-\lambda_{1\pm}^{2\mp}=\pm\Lambda\frac{h_1l_2-h_2l_1}{E_{{1\pm}}-E_{{2\mp}}}\,.
\end{align}
\end{subequations}
Note that the electric field mixes, e.g., $\ket{1-}$ with both $\ket{2-}$ and $\ket{2+}$. The above development holds only far from any (accidental) crossing between the $\ket{1\pm}$ and $\ket{2\pm}$ states.

\subsection{Effects of the static magnetic field}
\label{sec_magn}

The magnetic field Hamiltonian $H_{\rm m}=H_{\rm p}+H_{\rm d}+H_{\rm z}$ is the sum of three contributions. The first two ones, $H_{\rm p}$ and $H_{\rm d}$, result from the substitution $\vec{k}\rightarrow-i\vec{\nabla}+e\vec{A}/\hbar$ in the LK Hamiltonian, where $\vec{A}=-\vec{r}\times\vec{B}/2$ is the vector potential. The ``paramagnetic'' Hamiltonian $H_{\rm p}$ collects $\propto A_i$ terms while the ``diamagnetic'' Hamiltonian $H_{\rm d}$ collects $\propto A_iA_j$ terms ($i,j\in\{x,y,z\}$). The third contribution, $H_{\rm z}=2\kappa\mu_B\vec{B}\cdot\vec{J}$, is the Zeeman Hamiltonian that describes the action of the magnetic field onto the Bloch functions ($\vec{J}$ being the $3/2$ angular momentum of the holes).\cite{Luttinger56} 

As discussed in Ref. \onlinecite{Venitucci18}, the diamagnetic Hamiltonian $H_{\rm d}$ is not relevant for the calculation of the Larmor and Rabi frequencies of the qubit (to first order in the magnetic field) and will be dropped out in the following. The paramagnetic Hamiltonian $H_{\rm p}$ has no action in the minimal basis set $\cal{B}$. Hence, only the Zeeman Hamiltonian $H_{\rm z}$ has non-zero matrix elements in $\cal{B}$:
\begin{equation}
\bra{i, j_z}H_{\rm z}\ket{i^\prime, j_z^\prime}=\delta_{i, i^\prime}\bra{j_z}\mathcal{H}_{\rm z}\ket{j_z^\prime}\,,
\end{equation}
where $\{i,i^\prime\}\in\{1,2\}$, and $\mathcal{H}_{\rm z}$ reads in the $\{\ket{+\frac{3}{2}},\ket{+\frac{1}{2}},\ket{-\frac{1}{2}},\ket{-\frac{3}{2}}\big\}$ basis set:
\begin{equation}
\mathcal{H}_z=\kappa\mu_B B
\begin{pmatrix}
3b_z & \sqrt{3}b_-& 0 & 0 \\
\sqrt{3}b_+ & b_z & 2b_- & 0 \\
0 & 2b_+ & -b_z & \sqrt{3}b_- \\
0 & 0 & \sqrt{3}b_+ & -3b_z
\end{pmatrix}\,.
\label{eqHz}
\end{equation}
$\vec{b}=(b_x,b_y,b_z)$ the unit vector pointing along the magnetic field, and $b_+=b_-^*=b_x+i b_y$. At variance with the static electric field, which mixes $n_y=1$ and $n_y=2$ envelopes with the same angular momentum $j_z$ ($\Delta j_z=0$), the static magnetic field mixes non-orthogonal (same $n_y$) envelopes with different $j_z$'s ($\Delta j_z=\pm 1$).

\section{The Rabi frequency}
\label{sectionRabi}

We now compute the Rabi frequency of a qubit based on the hole states introduced in the previous section.

\subsection{General equations}
\label{subsectionRabigeneral}

We consider a qubit based on the ground hole states $\ket{\tilde{1}-,\Uparrow}$ and $\ket{\tilde{1}-,\Downarrow}$. These two states are degenerate at zero magnetic field but are split at finite $B$. \RED{In the following, we deal with the magnetic field using degenerate perturbation theory in order to reach a first-order, $\propto B$ expression for the Rabi frequency.} The zeroth-order qubit states $\ket{\zero_0}$ and $\ket{\one_0}$ and the first-order qubit energies $E_1(\zero)$ and $E_1(\one)$ are thus the eigensolutions of the Hamiltonian:\cite{Venitucci18} 
\begin{equation}
H_1(\vec{B})=
\begin{pmatrix}
\braket{\tilde{1}-,\Uparrow|H_{\rm m}^\prime|\tilde{1}-,\Uparrow} & \braket{\tilde{1}-,\Uparrow|H_{\rm m}^\prime|\tilde{1}-,\Downarrow} \\
\braket{\tilde{1}-,\Downarrow|H_{\rm m}^\prime|\tilde{1}-,\Uparrow} & \braket{\tilde{1}-,\Downarrow|H_{\rm m}^\prime|\tilde{1}-,\Downarrow}
\end{pmatrix}\,,
\label{eqH1}
\end{equation}
where $H_{\rm m}^\prime=H_{\rm p}+H_{\rm z}$ collects all $\propto B$ terms of the magnetic Hamiltonian $H_{\rm m}$. \RED{We emphasize that $H_{\rm p}$ has no action and $H_{\rm m}^\prime\equiv H_{\rm z}$ in the minimal basis set ${\cal B}$, yet not in the larger basis sets that will be considered in the numerical simulations of section \ref{sectionDiscussion}}. The same gates that apply the static electric field are used to drive Rabi oscillations between $\ket{\zero}$ and $\ket{\one}$ with a radio-frequency (RF) electric field modulation $\delta\vec{E}(t)=E_{\rm ac}\sin(2\pi f_L t+\phi)\vec{y}$ resonant with the Larmor frequency $f_L$ of the qubit. In these conditions, the Rabi frequency reads:
\begin{equation}
f_R=\frac{e}{h}E_{\rm ac}|\bra{\one}y\ket{\zero}|\,. 
\label{eqfr}
\end{equation}
As discussed in Ref. \onlinecite{Venitucci18}, $f_R$ can be computed to first order in $B$ from the first-order states:
\begin{subequations}
\label{eqzeroone1}
\begin{align}
\ket{\zero_1}&=\ket{\zero_0}+\sum_{n,\sigma}\frac{\bra{n,\sigma}H_{\rm m}^\prime\ket{\zero_0}}{E_{\tilde{1}-}-E_n}\ket{n,\sigma} \\
\ket{\one_1}&=\ket{\one_0}+\sum_{n,\sigma}\frac{\bra{n,\sigma}H_{\rm m}^\prime\ket{\one_0}}{E_{\tilde{1}-}-E_n}\ket{n,\sigma}\,,
\end{align}
\end{subequations}
where $\ket{n,\sigma}$ is any excited state with pseudo-spin $\sigma$. Substitution in Eq. (\ref{eqfr}) yields:
\begin{align}
f_R=
\frac{eE_{\rm ac}}{h}\Big|&\sum_{n,\sigma}\frac{1}{E_{\tilde{1}-}-E_n}\big(\bra{\one_0}y\ket{n,\sigma}\bra{n,\sigma}H_{\rm m}^\prime\ket{\zero_0} \nonumber \\
&+\bra{\one_0}H_{\rm m}^\prime\ket{n,\sigma}\bra{n,\sigma}y\ket{\zero_0}\big)\Big|\,.
\label{eqfrabisos}
\end{align}
 
We now aim to develop the Larmor and Rabi frequencies to first order in all fields, including $E_0$, in the basis set ${\cal B}$ \RED{where the sum over $n$ runs over $\ket{\tilde{1}+}$ and $\ket{\tilde{2}\pm}$. First of all, the energies $E_1(\zero)$ and $E_1(\one)$ and the states $\ket{\zero_0}$ and $\ket{\one_0}$ are, to first order in $E_0$, the eigensolutions of:}
\begin{equation}
H_1(\vec{B})=\frac{1}{2}\mu_BB
\begin{pmatrix}
g_z b_z & g_x b_x-ig_y b_y \\
g_x b_x+ig_y b_y & -g_z b_z 
\end{pmatrix}\,, 
\label{eqH12}
\end{equation}
where:
\begin{subequations}
\label{eqgs}
\begin{align}
g_x&=4\kappa\big(\sqrt{3}h_1l_1+l_1^2\big)\\
g_y&=4\kappa\big(\sqrt{3}h_1l_1-l_1^2\big)\\
g_z&=2\kappa\Big(3h_1^2-l_1^2\Big)\,.
\end{align}
\end{subequations}
The Larmor frequency is therefore:
\begin{equation}
f_L=\frac{1}{h}|E_1(\one)-E_1(\zero)|=\frac{\mu_BB}{h}\sqrt{g_x^2 b_x^2+g_y^2 b_y^2+g_z^2 b_z^2}\,.
\label{eqLarmor}
\end{equation}
$g_x$, $g_y$ and $g_z$ can be identified as the principal $g$-factors along the magnetic axes $x$, $y$ and $z$.\cite{Venitucci18} Also,
\begin{subequations}
\label{eqzeroone0}
\begin{align}
\ket{\zero_0}&={\alpha}\ket{\tilde{1}-,\Uparrow}+{\beta}\ket{\tilde{1}-,\Downarrow} \\
\ket{\one_0}&=-{\beta}\ket{\tilde{1}-,\Uparrow}+{\alpha}^*\ket{\tilde{1}-,\Downarrow}\,,
\end{align}
\end{subequations}
where:
% \begin{widetext}
\begin{subequations}
\label{eqalphabeta}
\begin{align}
\alpha&=\frac{-g_x b_x+ig_y b_y}{\sqrt{g_x^2 b_x^2+g_y^2 b_y^2+{\big(g_z b_z+\sqrt{g_x^2 b_x^2+g_y^2 b_y^2+g_z^2 b_z^2}\big)}^2}} \\
\beta&=\frac{g_z b_z+\sqrt{g_x^2 b_x^2+g_y^2 b_y^2+g_z^2 b_z^2}}{\sqrt{g_x^2 b_x^2+g_y^2 b_y^2+{\big(g_z b_z+\sqrt{g_x^2 b_x^2+g_y^2 b_y^2+g_z^2 b_z^2}\big)}^2}}\,.
\end{align}
\end{subequations}
% \end{widetext}

At zero static electric field, $E_{\rm ac}$ can only couple $\ket{1-}$ with $\ket{2-}$ and $\ket{2+}$ (through the dipole matrix elements along $y$), while $H_{\rm z}$ can only couple $\ket{1-}$ with $\ket{1+}$. Therefore, the Rabi frequency is zero as there are no excited states able to connect $\ket{\zero_0}$ and $\ket{\one_0}$ in Eq. (\ref{eqfrabisos}). This is supported by a symmetry analysis: When $E_0=0$, the system has three mirror planes perpendicular to $x$, $y$, $z$, which, as shown in Ref. \onlinecite{Venitucci18}, implies that the Rabi frequency is zero to first order in $B$ and $E_{\rm ac}$.

At first order in $E_0$, Eq. (\ref{eqfrabisos}) can hence be factorized as:
\begin{equation}
f_R=\frac{e}{h}B|E_0|E_{\rm ac}\Big|\Pi_{\tilde{1}+}+\Pi_{\tilde{2}-}+\Pi_{\tilde{2}+}\Big|\,,
\label{eqfrabipi}
\end{equation}
where $\Pi_{\tilde{1}+}$, $\Pi_{\tilde{2}+}$ and $\Pi_{\tilde{2}-}$ are the contributions of $\ket{\tilde{1}+}$, $\ket{\tilde{2}+}$ and $\ket{\tilde{2}-}$ to the sum-over-states and are given in Appendix \ref{AppendixSOS}.

Eq. (\ref{eqfrabipi}) together with Eqs. (\ref{eqipmnofield}), (\ref{eqtilde1}), (\ref{eqtilde2}), (\ref{eqzeroone0}) and Appendix \ref{AppendixSOS} provide an analytical model for $f_R$ to first order in all fields $B$, $E_0$ and $E_{\rm ac}$ in the minimal basis set $\mathcal{B}$. However, to make the expression of $f_R$ more tractable, we will further expand relevant quantities in powers of $L_z/L_x$ and $L_z/L_y$ in the ``thin dot'' limit $L_z\ll L_x, L_y$ suitable for most planar and SOI devices on Si $(001)$ substrates.

\subsection{The thin dot limit}

In the limit $L_z\ll L_x, L_y$, $\ket{1-}$ and $\ket{2-}$ are mostly heavy-hole states ($j_z=\pm 3/2$) while $\ket{1+}$ and $\ket{2+}$ are mostly light-hole states ($j_z=\pm 1/2$). Therefore, $E_0$ and $E_{\rm ac}$ essentially couple $\ket{1-,\sigma}$ and $\ket{2-,\sigma}$. To lowest orders in $L_z/L_x$ and $L_z/L_y$, only $\ket{\tilde{2}-}$ states actually make a contribution to the Rabi frequency in Eq. (\ref{eqfrabisos}). More specifically, the $\ket{1-,\Uparrow}$ and $\ket{2-,\Uparrow}$ states [Eqs. (\ref{eqipmnofield})] read to second order in $L_z/L_x$ and $L_z/L_y$:
\begin{subequations}
\begin{align}
\ket{1-,\Uparrow}&=\ket{1,+\frac{3}{2}}+\delta l_1\ket{1,-\frac{1}{2}} \\
\ket{2-,\Uparrow}&=\ket{2,+\frac{3}{2}}+\delta l_2\ket{2,-\frac{1}{2}}\,,
\end{align}
\end{subequations}
where:
\begin{subequations}
\label{eqdeltal}
\begin{align}
\delta l_1&=-\frac{\sqrt{3}}{4}\frac{\gamma_3}{\gamma_2}\Big(\frac{L_z^2}{L_y^2}-\frac{L_z^2}{L_x^2}\Big) \\
\delta l_2&=-\frac{\sqrt{3}}{4}\frac{\gamma_3}{\gamma_2}\Big(4\frac{L_z^2}{L_y^2}-\frac{L_z^2}{L_x^2}\Big)\,.
\end{align}
\end{subequations}
Next, at finite $E_0$,
\begin{subequations}
\begin{align}
\ket{\tilde{1}-,\Uparrow}&=\ket{1,+\frac{3}{2}}+\lambda\ket{2,+\frac{3}{2}} \nonumber \\
&+\delta l_1\ket{1,-\frac{1}{2}}+\lambda\delta l_2\ket{2,-\frac{1}{2}} \\
\ket{\tilde{2}-,\Uparrow}&=\ket{2,+\frac{3}{2}}-\lambda\ket{1,+\frac{3}{2}} \nonumber \\
&+\delta l_2\ket{2,-\frac{1}{2}}-\lambda\delta l_1\ket{1,-\frac{1}{2}}\,,
\end{align}
\end{subequations}
where [see Eqs. (\ref{eqtilde1})]:
\begin{equation}
\lambda=-\frac{32m_0eE_0L_y^3}{27\pi^4\hbar^2(\gamma_1+\gamma_2)}\,.
\label{eqlambda}
\end{equation}
The expressions are similar for $\ket{\tilde{1}-,\Downarrow}$ and $\ket{\tilde{2}-,\Downarrow}$ (with $j_z=3/2$ replaced by $j_z=-3/2$ and $j_z=-1/2$ by $j_z=1/2$). 

The principal $g$-factors [Eqs. (\ref{eqgs})] are then, to second order in $L_z/L_x$ and $L_z/L_y$:
\begin{subequations}
\label{eqgHH}
\begin{align}
g_x&=g_y=4\sqrt{3}\kappa\delta l_1=-3\kappa\frac{\gamma_3}{\gamma_2}\Big(\frac{L_z^2}{L_y^2}-\frac{L_z^2}{L_x^2}\Big)\\
g_z&=6\kappa\,.
\end{align}
\end{subequations}
As expected for mostly heavy-hole states, $|g_z|\gg |g_x|, |g_y|$.

\begin{figure}
\includegraphics[width=0.95\columnwidth]{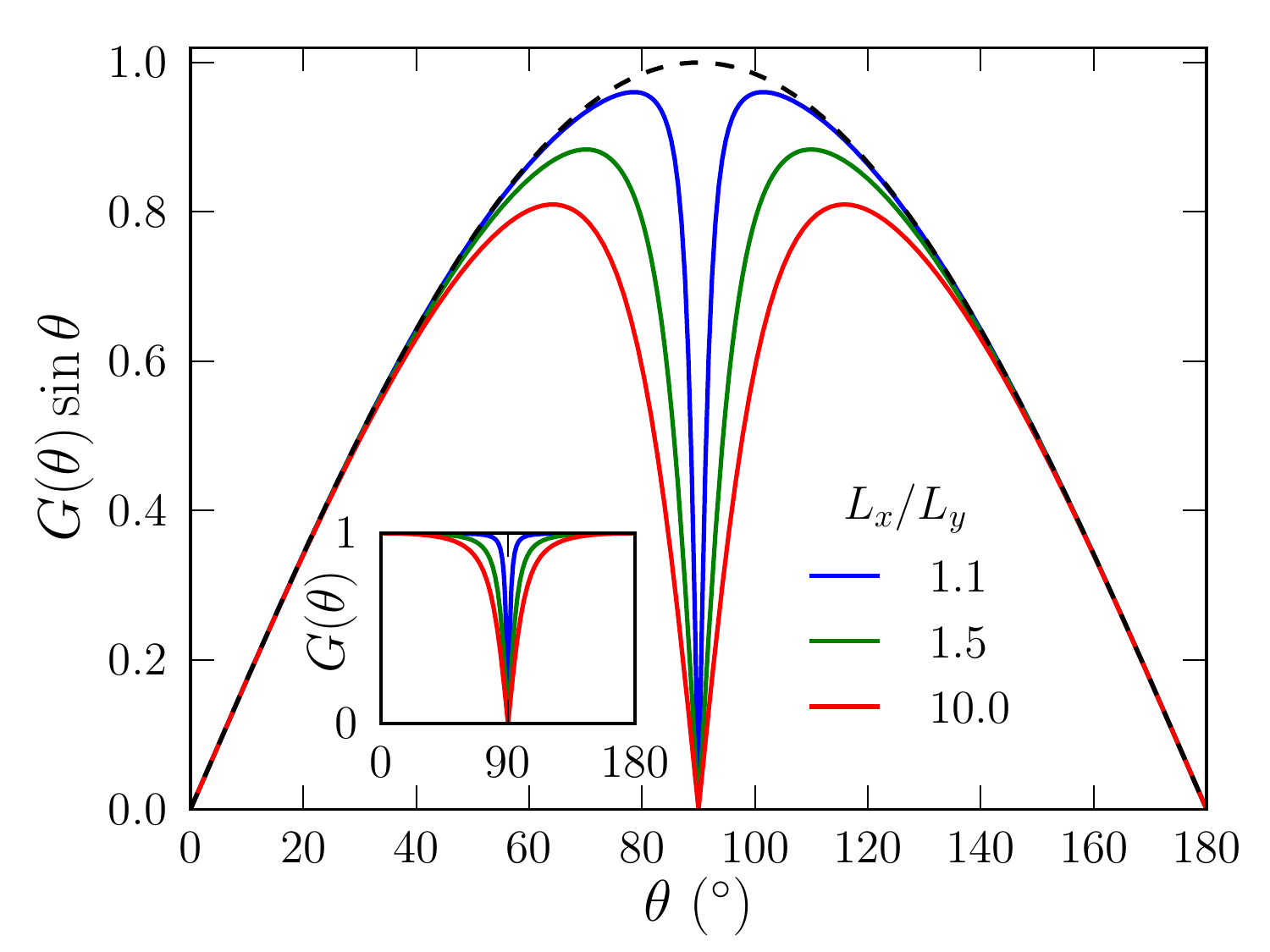}
\caption{The function $G(\theta)\sin\theta$ for different ratios $L_x/L_y$ at $L_z/L_y=1/3$. The dashed black line is the $\sin\theta$ envelope.}
\label{figH}
\end{figure}

In Eq. (\ref{eqfrabisos}), the matrix elements of $H_{\rm z}$ between states $\{\ket{\tilde{1}-,\Uparrow}, \ket{\tilde{1}-,\Downarrow}\}$ (columns) and $\{\ket{\tilde{2}-,\Uparrow}, \ket{\tilde{2}-,\Downarrow}\}$ (rows) read, to second order in $L_z/L_x$ and $L_z/L_y$:
\begin{equation}
\mathcal{H}_{\rm z}^{(21)}=2\sqrt{3}\kappa\mu_B B\lambda(\delta l_2-\delta l_1)
\begin{pmatrix}
0 & b_- \\
b_+ & 0
\end{pmatrix}\,.
\label{eqHz21}
\end{equation}
As a matter of explanation, the matrix elements between opposite pseudo-spins result from the $\propto b_\pm$ interaction of the majority $j_z=\pm 3/2$ component of one pseudo-spin with the minority $j_z=\pm 1/2$ component of the other. These two envelopes are orthogonal if $E_0=0$ or $\delta l_1=\delta l_2$ and can not, therefore, be coupled by $H_{\rm z}$ (Indeed, $\ket{\tilde{1}-,\Uparrow}$ and $\ket{\tilde{2}-,\Uparrow}$ can then be factorized as the products of single, orthogonal envelopes by the same mixed heavy- and light-hole Bloch function). This gives rise to the $\propto\lambda(\delta l_2-\delta l_1)$ dependence in Eq. (\ref{eqHz21}). The physics of the Rabi oscillations will be further analyzed in section \ref{subsectionPhysics}. Substituting the above equations into the expression for the Rabi frequency yields:
\begin{equation}
f_R=\frac{64\sqrt{3}e}{9\pi^2h}\mu_B|\kappa|BE_{\rm ac}L_y\frac{|\lambda|(\delta l_2-\delta l_1)}{E_{2-}-E_{1-}}\big|\alpha^2b_+-\beta^2b_-\big|\,.
\label{eqfr2t}
\end{equation}
The last term rules the dependence of the Rabi frequency on the orientation of the magnetic field, and can be factorized as:
\begin{equation}
|\alpha^2b_+-\beta^2b_-\big|=G(\theta)\sin\theta\,,
\end{equation}
where $\theta$ is the polar angle between the magnetic field and the $z$ axis (see Fig. \ref{figqubit}), and:
\begin{subequations}
\begin{align}
G(\theta)&=\frac{1}{\sqrt{1+F^2(\theta)}} \\
F(\theta)&=\frac{\gamma_3}{2\gamma_2}\Big(\frac{L_z^2}{L_y^2}-\frac{L_z^2}{L_x^2}\Big)\tan\theta\,.
\end{align}
\end{subequations}
The function $F(\theta)$ has been expanded to second order in $L_z/L_x$ and $L_z/L_y$. It is, however, practically not relevant to expand $G(\theta)$ in powers of $L_z/L_x$ and $L_z/L_y$ as the convergence of the resulting series is highly non uniform with respect to the variable $\theta$. Expanding only the prefactor of $G(\theta)$ in Eq. (\ref{eqfr2t}) to second order in $L_z/L_x$ and $L_z/L_y$ finally yields:
\begin{equation}
f_R^{(2)}=\frac{2^8m_0e^3}{3^4\pi^9\hbar^4}B|E_0|E_{\rm ac}\frac{\gamma_3|\kappa|}{\gamma_2(\gamma_1+\gamma_2)^2}L_y^6\frac{L_z^2}{L_y^2}G(\theta)\sin\theta\,,
\label{eqfr2}
\end{equation}
At this level of approximation, the Rabi frequency does not depend on the azimuthal angle $\varphi$. The $\sin\theta$ envelope results from the $\propto b_\pm$ dependence of the matrix elements of $H_z$ [Eq. (\ref{eqHz21})]. The function $G(\theta)$ arises from the interplay with the pseudo-spin composition of $\ket{\zero_0}$ and $\ket{\one_0}$ [the $\alpha$ and $\beta$ coefficients in Eq. (\ref{eqfr2t})]. $G(\theta)\sim 1$ everywhere except near $\theta=\pi/2$ where it shows a dip (see Fig. \ref{figH}), whose origin will be discussed later. 

The Rabi frequency gets dependent on $\varphi$ once $\ket{\tilde{1}-}$ and $\ket{\tilde{2}-}$ are expanded to fourth order in $L_z/L_x$ and $L_z/L_y$. After further algebra,
\begin{align}
f_R^{(4)}&=f_R^{(2)}\Big\{1+\frac{1}{4\gamma_2(\gamma_1+\gamma_2)}\times \nonumber \\
&\Big[A_1\frac{L_z^2}{L_y^2}-A_2\frac{L_z^2}{L_x^2}+A_3\Big(5\frac{L_z^2}{L_y^2}-2\frac{L_z^2}{L_x^2}\Big)\cos2\varphi\Big]\Big\}\,, 
\label{eqfr4}
\end{align}
where $A_1=10(\gamma_1 \gamma_2+\gamma_2^2+3 \gamma_3^2)$, $A_2=12\gamma_3^2$, and $A_3=\gamma_3(\gamma_1+\gamma_2)$. 

\subsection{High field electric corrections}

Eqs. (\ref{eqfrabipi}), (\ref{eqfr2}) and (\ref{eqfr4}) are valid at small static electric field $E_0$. However, as shown in the next section, the Rabi frequency decreases at large $E_0$, in particular because the dipole matrix element $\bra{\tilde{2}-}y\ket{\tilde{1}-}$ dies out once the $\ket{\tilde{1}-}$ and $\ket{\tilde{2}-}$ states get spatially separated by the static electric field.\cite{Venitucci18}

An expression for the Rabi frequency accurate for arbitrary electric fields can be derived when $\lambda_{2\mp}^{1\pm}$ and $\lambda_{1\pm}^{2\mp}$ are negligible [Eqs. (\ref{eqlambdas})]. In that limit, the electric field couples $\ket{1-}$ to $\ket{2-}$ but not to $\ket{2+}$; the resulting two-level Hamiltonian can be solved exactly for the eigenstates $\ket{\tilde{1}-}$ and $\ket{\tilde{2}-}$. Keeping track of the exact expression of $E_{\tilde{1}-}$, $\ket{\tilde{1}-}$, $E_{\tilde{2}-}$ and $\ket{\tilde{2}-}$ everywhere except in $\alpha$ and $\beta$, the Rabi frequency [Eq. (\ref{eqfrabipi})] simply gets renormalized by a factor:
\begin{equation}
F_e(E_0)=\Big[1+\frac{1}{2}\Big(\frac{E_0}{E_{\rm max}}\Big)^2\Big]^{-\frac{3}{2}}\,,
\label{eqFe}
\end{equation}
where:
\begin{equation}
E_{\rm max}^{-1}=2\sqrt{2}e\frac{\big|\bra{2-}y\ket{1-}\big|}{E_{2-}-E_{1-}}\,.
\end{equation}
This approximation is relevant in the thin dot limit. To lowest order in $L_z/L_x$ and $L_z/L_y$, $E_{\rm max}$ then reads:
\begin{equation}
E_{\rm max}^{(0)}=\frac{27\pi^4\hbar^2(\gamma_1+\gamma_2)}{64\sqrt{2}m_0eL_y^3}\,.
\label{eqEmax0}
\end{equation}
The renormalized Rabi frequencies $\tilde{f}_R^{(2)}(E_0)=f_R^{(2)}(E_0)F_e(E_0)$ and $\tilde{f}_R^{(4)}(E_0)=f_R^{(4)}(E_0)F_e(E_0)$ are maximum when $E_0=E_{\rm max}$. At this field, $\tilde{f}_R(E_{\rm max})=f_R(\tilde{E}_{\rm max})$, where $\tilde{E}_{\rm max}=(3/2)^{-3/2}E_{\rm max}$. In particular,
\begin{equation}
\tilde{f}_R^{(2)}(E_{\rm max}^{(0)})=\frac{8e^2}{9\sqrt{3}\pi^5\hbar^2}BE_{\rm ac}\frac{\gamma_3|\kappa|}{\gamma_2(\gamma_1+\gamma_2)}L_y^3\frac{L_z^2}{L_y^2}G(\theta)\sin\theta\,. 
\end{equation}
Note that the maximal Rabi frequency scales as $L_yL_z^2$ in the thin dot limit $L_z\ll L_y$.

\section{Discussion}
\label{sectionDiscussion}

\subsection{Validation of the model}

In order to test the above model and approximations, we consider a silicon box with sides $L_x=40$ nm, $L_y=30$ nm and $L_z=10$ nm subjected to a static magnetic field $B=1$ T parallel to $\vec{y}+\vec{z}$ ($\theta=45^\circ$, $\varphi=0^\circ$). The RF electric field is $E_{\rm ac}=0.03$ mV/nm. The material parameters $\kappa$, $\gamma_1$, $\gamma_2$ and $\gamma_3$ are given in Table \ref{tabmat}. 

\begin{figure}
\includegraphics[width=0.95\columnwidth]{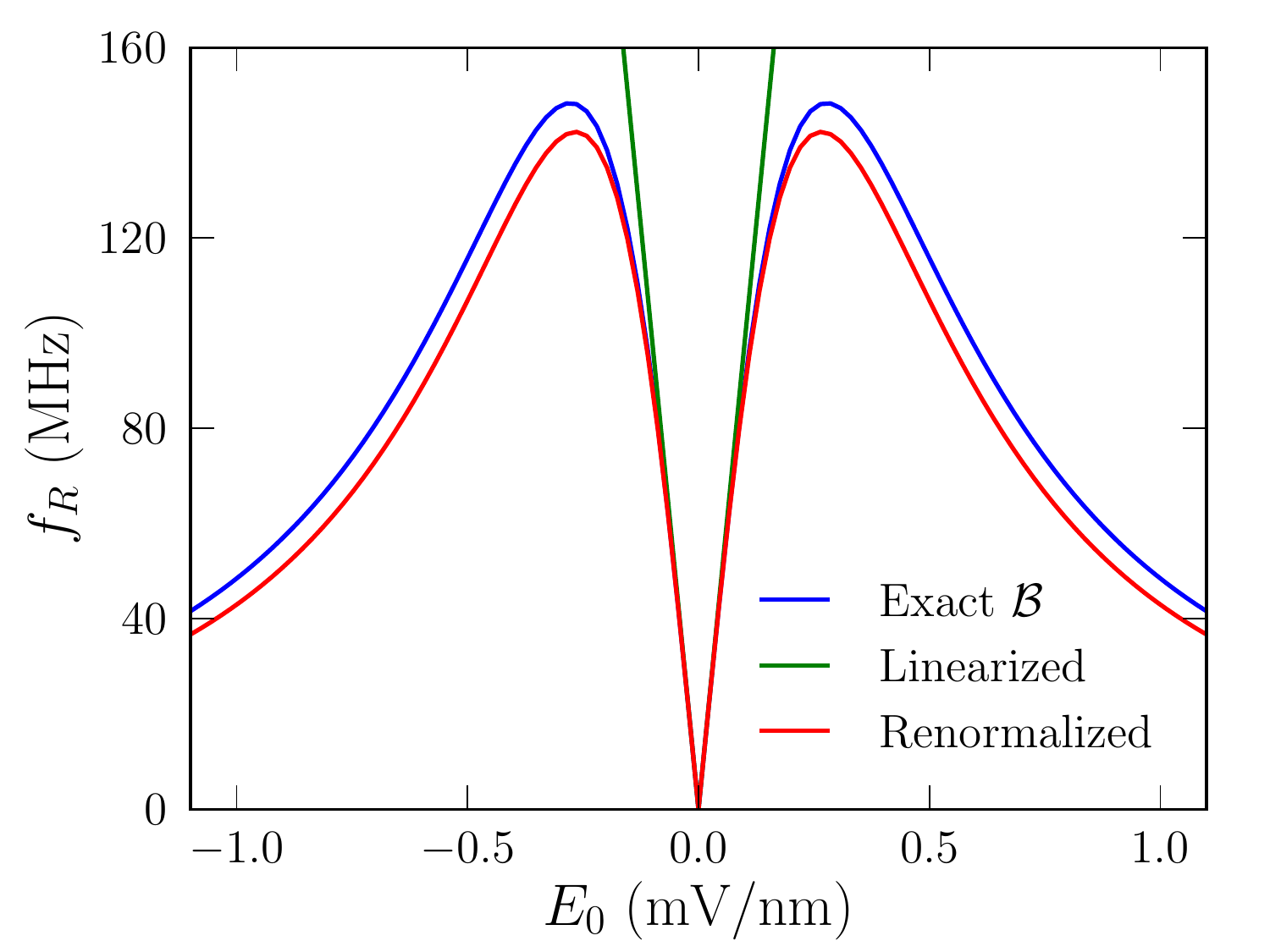}
\caption{Rabi frequency as a function of the static electric field $E_0$ in a silicon quantum dot with sides $L_x=40$ nm, $L_y=30$ nm and $L_z=10$ nm. The magnetic field $B=1$ T is oriented along $\vec{y}+\vec{z}$. The RF electric field is $E_{\rm ac}=0.03$ mV/nm. The Rabi frequency is computed either from Eq. (\ref{eqfrabisos}) using the exact eigenstates of the Hamiltonian in the minimal basis set $\cal{B}$ [``Exact $\cal{B}$''], or from the approximation to first order in $E_0$ [``Linearized'', Eq. (\ref{eqfrabipi})], then renormalized by Eq. (\ref{eqFe}) [``Renormalized''].}
\label{figfrE0}
\end{figure}

\begin{table}
\RED{
\centering
\begin{tabular}{|l|d|d|d|d|d|d|}
\hline
& \multicolumn{1}{c|}{Si} & \multicolumn{1}{c|}{Ge} & \multicolumn{1}{c|}{InP} & \multicolumn{1}{c|}{GaAs} & \multicolumn{1}{c|}{InAs} & \multicolumn{1}{c|}{InSb} \\ 
\hline
$E_g$ (eV) & 4.34 & 0.89 & 1.42 & 1.52 & 0.42 & 0.24 \\
$\Delta$ (eV) & 0.044 & 0.29 & 0.11 & 0.34 & 0.41 & 0.80 \\
$\gamma_1$ & 4.285 & 13.38 & 4.95 & 6.85 & 20.40 & 37.10 \\
$\gamma_2$ & 0.339 & 4.24 & 1.65 & 2.10 & 8.30 & 16.50 \\
$\gamma_3$ & 1.446 & 5.69 & 2.35 & 2.90 & 9.10 & 17.70 \\
$m_{z}$ ($m_0$) & 0.277 & \multicolumn{1}{r|}{0.204} & \multicolumn{1}{r|}{0.606} & \multicolumn{1}{r|}{0.377} & \multicolumn{1}{r|}{0.263} & \multicolumn{1}{r|}{0.244} \\
$m_{xy}$ ($m_0$) & 0.216 & \multicolumn{1}{r|}{0.057} & \multicolumn{1}{r|}{0.152} & \multicolumn{1}{r|}{0.112} & \multicolumn{1}{r|}{0.035} & \multicolumn{1}{r|}{0.019} \\
$\kappa$   & -0.42 & 3.41 & 0.97 & 1.20 & 7.60 & 15.60 \\
\hline
$\zeta_{[110]}$ ($\times 100$) & 8.38 & 1.47 & 3.17 & 2.07 & 1.01 & 0.58 \\
$\zeta_{[001]}$ ($\times 100$) & 1.96 & 1.10 & 2.23 & 1.50 & 0.92 & 0.54 \\
\hline
$\zeta_{[110]}^\prime$ ($\times 100$) & 92.25 & 7.62 & 21.58 & 15.43 & 3.82 & 2.00 \\
$\zeta_{[001]}^\prime$ ($\times 100$) & 21.63 & 5.68 & 15.15 & 11.17 & 3.48 & 1.87 \\
\hline
\end{tabular}}
\caption{Bandgap energy $E_g$ at $\Gamma$, spin-orbit splitting energy $\Delta$ in the valence band, Luttinger parameters \RED{and masses of the heavy-holes along $z$ [$m_{z}=m_0/(\gamma_1-2\gamma_2)$], and in the $(xy)$ plane [$m_{xy}=m_0/(\gamma_1+\gamma_2)$]}, $\kappa$ parameter\cite{Winkler03} and coefficients $\zeta_{[110]}$, $\zeta_{[100]}$, $\zeta_{[110]}^\prime$ and $\zeta_{[100]}^\prime$ characterizing the speed of Rabi oscillations in $[110]$- and $[100]$-oriented dots, for different materials [Eqs. (\ref{eqzeta110}), (\ref{eqzeta100}) and (\ref{eqzetap})].}
\label{tabmat}
\end{table}

The Rabi frequency is plotted as a function of the static electric field $E_0$ in Fig. \ref{figfrE0}. It is computed either from Eq. (\ref{eqfrabisos}), using the exact eigenstates of the Hamiltonian in the minimal basis set $\cal{B}$ as inputs, or from the approximation to first order in $E_0$ [Eq. (\ref{eqfrabipi})]. The thin dot limit is not taken at this stage. As expected, the first-order approximation reproduces the slope of the Rabi frequency around $E_0=0$. However, at larger field the Rabi frequency computed from the exact eigenstates drops owing to the decrease of the dipole matrix elements $\bra{\tilde{2}-}y\ket{\tilde{1}-}$ and to the increase of $E_{\tilde{2}-}-E_{\tilde{1}-}$ in the steep triangular well created by the static electric field.\cite{Venitucci18} This trend is, nonetheless, very well captured by the renormalized first-order approximation [Eq. (\ref{eqFe})].

\begin{figure}
\includegraphics[width=0.95\columnwidth]{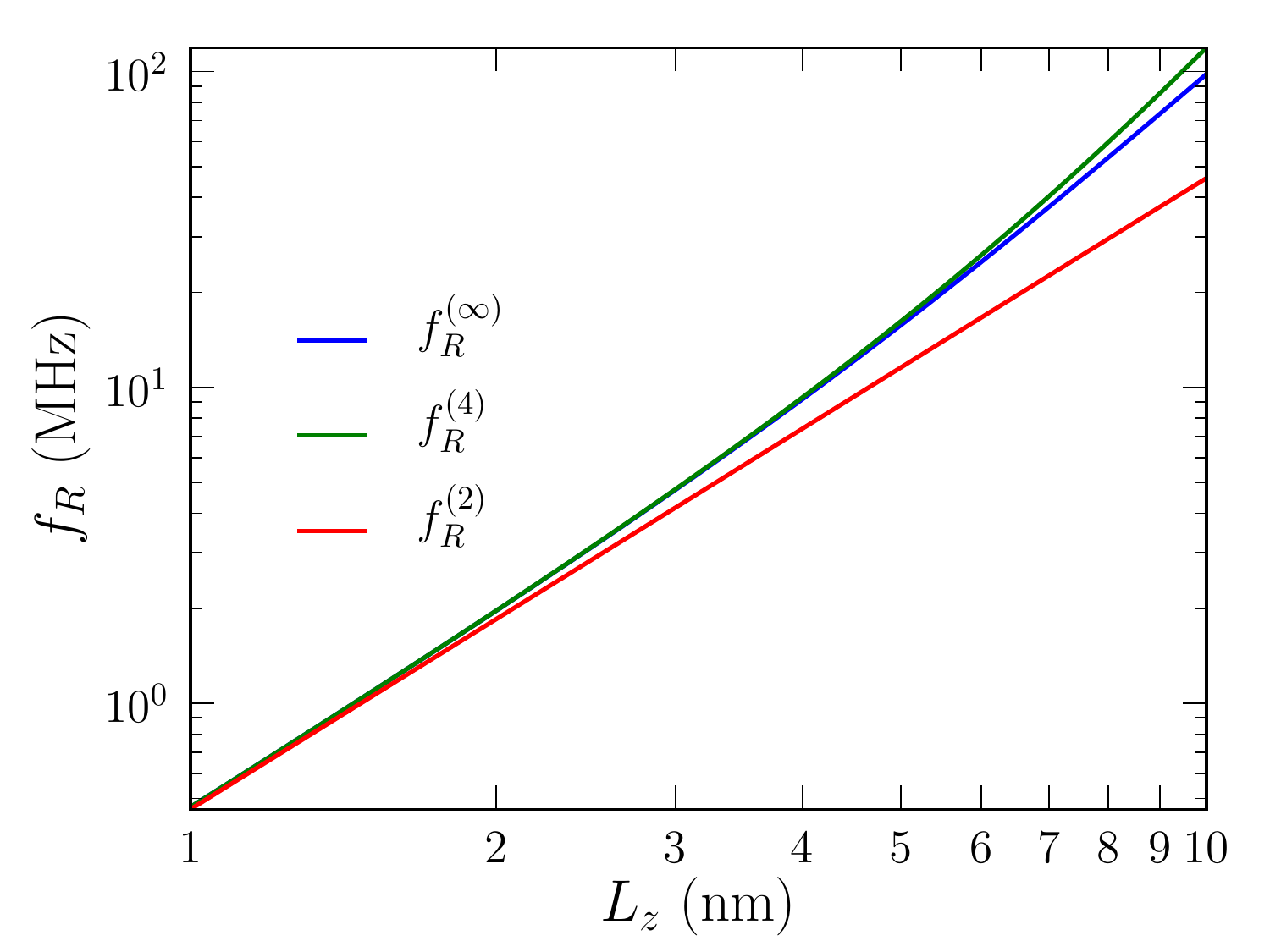}
\caption{Rabi frequency as a function of the height of the dot $L_z$ in three different approximations, $f_R^{(2)}$ [Eq. (\ref{eqfr2})], $f_R^{(4)}$ [Eq. (\ref{eqfr4})], and $f_R^{(\infty)}$ [Eq. (\ref{eqfrabipi})], all linearized with respect to the static field $E_0$. Here $L_x=40$ nm, $L_y=30$ nm, $E_0=0.1$ mV/nm, $E_{\rm ac}=0.03$ mV/nm, and $B=1$ T parallel to $\vec{y}+\vec{z}$.}
\label{figfrLz}
\end{figure}

We next compare the expansions to order $L_z^2$ ($f_R^{(2)}$) and $L_z^4$ ($f_R^{(4)}$) to the ``all-orders'' $f_R^{(\infty)}$ defined by Eq. (\ref{eqfrabipi}). The three Rabi frequencies (which are all linearized with respect to $E_0$) are plotted in Fig. \ref{figfrLz} as a function of $L_z$ (for the same $L_x$ and $L_y$ as before), at $E_0=0.1$ mV/nm. While $f_R^{(2)}$ can be significantly smaller than $f_R^{(\infty)}$ , $f_R^{(4)}$ is much closer (but always slightly larger) in the whole $L_z=1-10$ nm range.

\begin{figure*}
\includegraphics[width=0.90\textwidth]{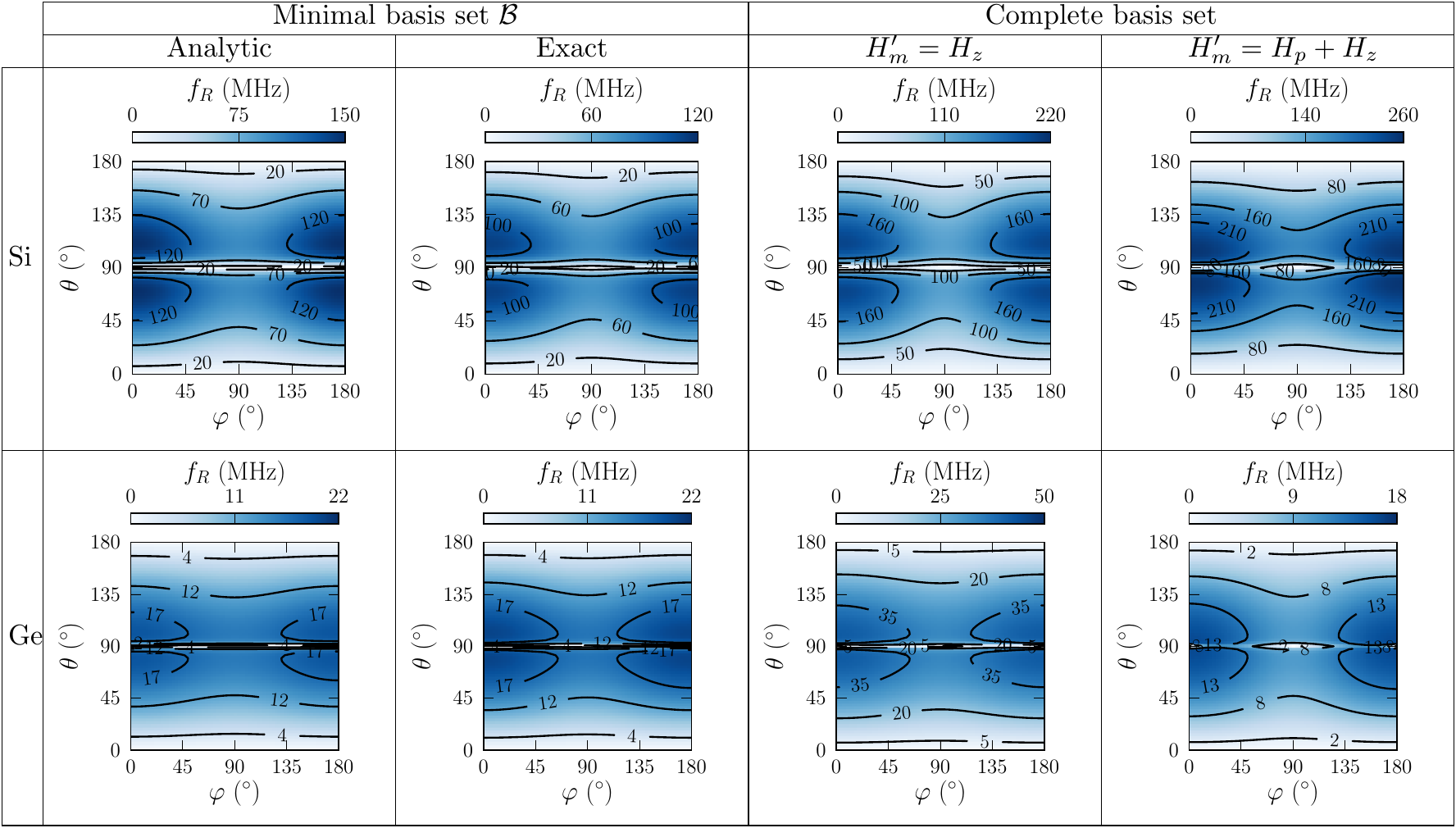}
\caption{Maps of Rabi frequency as a function of the orientation of the magnetic field for silicon (first row) and germanium (second row) dots with sides $L_x=40$ nm, $L_y=30$ nm and $L_z=10$ nm. The fields are $E_0=0.1$ mV/nm, $E_{\rm ac}=0.03$ mV/nm, and $B=1$ T. Four different approximations to the Rabi frequency are compared: (first column) fourth-order analytical formula [Eq. (\ref{eqfr4})]; (second column) exact solution of the model in the minimal basis set $\cal{B}$; (third column) exact solution in a ``converged'' basis set taking only the Zeeman Hamiltonian into account; (fourth column) exact solution in the same basis set accounting for the action of the vector potential on the envelope functions.}
\label{figfrmaps}
\end{figure*}

Finally, we compare in Fig. \ref{figfrmaps} the dependence of the Rabi frequency on the orientation of the magnetic field for two materials (Si, Ge) and four different approximations: {\it i}) the fourth-order analytical formula [Eq. (\ref{eqfr4})], {\it ii}) the exact solution of the model in the minimal basis set $\cal{B}$, {\it iii}) the exact solution in a ``converged'' basis set including \RED{quantum numbers} up to $n_x=n_y=n_z=18$, but taking only the Zeeman Hamiltonian into account ($H_{\rm m}^\prime=H_{\rm z}$), and {\it iv}) the exact solution in the same basis set now accounting for the action of the vector potential on the envelope functions ($H_{\rm m}^\prime=H_{\rm z}+H_{\rm p}$). The sides of the box are the same as in Fig. \ref{figfrE0}, and the material parameters for germanium are also given in Table \ref{tabmat}. The static electric field is $E_0=0.1$ mV/nm and the magnitude of the magnetic field is $B=1$ T. The analytical formula, Eq. (\ref{eqfr4}), provides a reasonable description of the orientational dependence of the Rabi frequency, which is, moreover, consistent with the maps computed in realistic SOI devices (going beyond the present simple box model) in Ref. \onlinecite{Venitucci18}. The Rabi frequency from the exact solution of the model in the minimal basis set $\cal{B}$ is only slightly different due to higher-order corrections to Eq. (\ref{eqfr4}). $f_R$ significantly increases in a larger basis set that picks the contributions from higher excited states, but still shows the same anisotropy. In that case, the paramagnetic Hamiltonian $H_{\rm p}$ (which has no action in $\cal{B}$) makes a sizable correction to the Rabi frequency. While the contributions from the Zeeman Hamiltonian are proportional to $\kappa$, those of the paramagnetic Hamiltonian scale as $\gamma_2$ and $\gamma_3$. They are actually opposite for Si (where $H_{\rm p}$ increases the Rabi frequency) and Ge (where $H_{\rm p}$ decreases it) owing to the opposite sign of $\kappa$ in the two materials. 

We conclude from the above discussion that the analytical formulas for $f_R^{(2)}$ and $f_R^{(4)}$ [Eqs. (\ref{eqfr2}) and Eqs. (\ref{eqfr4})] provide a semi-quantitative description of the Rabi oscillations and can be used to analyze the underlying physics as well as to outline trends \RED{in quantum dot material and geometry}.

\subsection{Physics of the Rabi oscillations}
\label{subsectionPhysics}

In this section, we discuss in more detail the physics behind Eq. (\ref{eqfr4}), and in particular its dependence on the dimensions of the box and Luttinger parameters.

According to Eq. (\ref{eqfr4}), the Rabi frequency primarily scales (in the thin dot limit) as $\zeta_{[110]}L_y^6(L_z^2/L_y^2)$, where:
\begin{equation}
\zeta_{[110]}=\frac{\gamma_3|\kappa|}{\gamma_2(\gamma_1+\gamma_2)^2}\,.
\label{eqzeta110}
\end{equation}
The $[110]$ subscript labels the orientation of the electric field (see next section for a discussion on box orientation). This equation highlights the ingredients needed to achieve Rabi oscillations driven by direct spin-orbit interactions in the valence band. 

First of all, there must be significant heavy- and light-hole mixing owing to lateral confinement in the hole ground-states and/or in the relevant excited states.\cite{Kloeffel18} Indeed, if all states are either pure $j_z=\pm 3/2$ or pure $j_z=\pm 1/2$ envelopes at $B=0$, then the qubit states $\ket{\zero_0}$ and $\ket{\one_0}$ have the same envelope function but time-reversal symmetric heavy-hole Bloch functions. The RF electric field $E_{\rm ac}$ then couples $\ket{\zero_0}$ and $\ket{\one_0}$ to heavy-hole \RED{excited states} with the same Bloch function but orthogonal envelopes. The Zeeman Hamiltonian $H_{\rm z}$ can not, however, couple orthogonal envelopes. The paramagnetic Hamiltonian $H_{\rm p}$ is not able to mix pure heavy-hole envelopes either (even in larger basis sets). As a consequence, there are no excited states able to connect $\ket{\zero_0}$ and $\ket{\one_0}$ in Eq. (\ref{eqfrabisos}). The argument also holds for light-hole qubit states. Therefore, there must be some degree of heavy- and light-hole mixing in the qubit or excited states at $B=0$. However, the static electric field can not mix heavy- and light-hole envelopes if such mixing does not pre-exist at $E_0=0$ (because it is diagonal in $j_z$). Hence lateral confinement is the primary driving force for the heavy- and light-hole mixing that is necessary to sustain electrically-driven Rabi oscillations. 

In the minimal basis set $\cal{B}$, the coupling of $j_z=\pm 3/2$ and $j_z=\pm 1/2$ envelopes by lateral confinement is characterized by $R_1$ and $R_2$, and is hence proportional to $\gamma_3$. To lowest order in perturbation, the resulting mixing between the heavy- and light-hole envelopes is inversely proportional to the splitting $2Q_i\propto\gamma_2$ between pure heavy- and light-hole states [see Eqs. (\ref{eqdeltal}) for the expressions of the light-hole mixings $\delta l_1$ and $\delta l_2$ in the thin dot limit]. This explains the $\gamma_3/\gamma_2$ factor in $\zeta_{[110]}$: the larger the coupling between between heavy- and light-holes with respect to their splitting ($\gamma_3\gg\gamma_2$), the faster the Rabi oscillations. A more careful analysis shows that there must actually be an imbalance between the heavy- and light-hole mixing in the ground $\ket{1-}$ and excited state $\ket{2-}$ [see discussion after Eq. (\ref{eqHz21})]: this is why the Rabi frequency is proportional to $\delta l_2-\delta l_1\propto(\gamma_3/\gamma_2)(L_z^2/L_y^2)$ in Eq. (\ref{eqfr2}).

The heavy- and light-hole mixing by lateral confinement is not, however, sufficient to allow for electrically-driven Rabi oscillations. Indeed, the RF electric field $E_{\rm ac}$ can not couple envelopes with same parities (with respect to the center of the box).\cite{Venitucci18} Yet lateral confinement mixes heavy- and light-hole envelopes with the same parity, and so do the Zeeman Hamiltonian $H_{\rm z}$ and the paramagnetic Hamiltonian $H_{\rm p}$ in Eqs. (\ref{eqzeroone1}). Only the static electric field $E_0$ does mix odd $p_y$ envelopes into the even $s$-like qubit ground-state, and is, therefore, an other pre-requisite for the Rabi oscillations. The mixing is actually proportional to $\Lambda\propto L_y$ and inversely proportional to the splitting $\Delta E\propto(\gamma_1+\gamma_2)/L_y^2$ between the $\ket{1-}\approx\ket{1,\pm\frac{3}{2}}$ and $\ket{2-}\approx\ket{2,\pm\frac{3}{2}}$ states, hence proportional to $L_y^3/(\gamma_1+\gamma_2)$.

Finally, $\ket{\tilde{1}-}$ and $\ket{\tilde{2}-}$ states are coupled by the Zeeman Hamiltonian in Eq. (\ref{eqzeroone1}). The mixing is proportional to $\kappa$ and, again, inversely proportional to the splitting $\Delta E\propto(\gamma_1+\gamma_2)/L_y^2$ between the $\ket{\tilde{1}-}\approx\ket{1-}$ and $\ket{\tilde{2}-}\approx\ket{2-}$ states. It breaks time-reversal symmetry in Eqs. (\ref{eqzeroone1}), which enables electrically-driven Rabi oscillations between $\ket{\zero_1}$ and $\ket{\one_1}$. The coupling to the RF electric field $E_{\rm ac}$ being proportional to $L_y$ (as is the coupling to $E_0$), the Rabi frequency scales altogether as $\gamma_3|\kappa|/[\gamma_2(\gamma_1+\gamma_2)^2]\times L_y^4/L_z^2$.

The orientational dependence of the Rabi frequency results from the interplay between the magnetic response of the heavy- and light-hole components. Indeed, we may define:
\begin{subequations}
\label{eqprimespin1}
\begin{align}
\ket{\tilde{1}-,\Downarrow^\prime}&\equiv\ket{\zero_0}={\alpha}\ket{\tilde{1}-,\Uparrow}+{\beta}\ket{\tilde{1}-,\Downarrow} \\
\ket{\tilde{1}-,\Uparrow^\prime}&\equiv\ket{\one_0}=-{\beta}\ket{\tilde{1}-,\Uparrow}+{\alpha}^*\ket{\tilde{1}-,\Downarrow}\,,
\end{align}
\end{subequations}
and apply the same transformation to $\ket{\tilde{2}-}$:
\begin{subequations}
\label{eqprimespin2}
\begin{align}
\ket{\tilde{2}-,\Downarrow^\prime}&={\alpha}\ket{\tilde{2}-,\Uparrow}+{\beta}\ket{\tilde{2}-,\Downarrow} \\
\ket{\tilde{2}-,\Uparrow^\prime}&=-{\beta}\ket{\tilde{2}-,\Uparrow}+{\alpha}^*\ket{\tilde{2}-,\Downarrow}\,,
\end{align}
\end{subequations}
where $\alpha$ and $\beta$ are given by Eqs. (\ref{eqalphabeta}). The RF electric field couples $\ket{\tilde{1}-,\Uparrow^\prime}$ to $\ket{\tilde{2}-,\Uparrow^\prime}$ and $\ket{\tilde{1}-,\Downarrow^\prime}$ to $\ket{\tilde{2}-,\Downarrow^\prime}$. In order to allow for Rabi oscillations, $H_{\rm z}$ must hence be able to mix $\ket{\tilde{2}-,\Downarrow^\prime}$ into $\ket{\tilde{1}-,\Uparrow^\prime}$, and $\ket{\tilde{2}-,\Uparrow^\prime}$ into $\ket{\tilde{1}-,\Downarrow^\prime}$. When there is a significant $b_z$ component, $\ket{\Uparrow^\prime}\approx\ket{\Uparrow}$ and $\ket{\Downarrow^\prime}\approx\ket{\Downarrow}$ as $|g_z|\gg |g_x|, |g_y|$ for mostly heavy-hole states [Eqs. (\ref{eqgHH})]. This large $g_z\simeq 6\kappa$ is the fingerprint of the strong, $\propto b_z$ splitting between the majority $\ket{\pm 3/2}$ components of $\ket{\tilde{1}-}$. The coupling $\bra{\tilde{2}-,\Downarrow}H_{\rm z}\ket{\tilde{1}-,\Uparrow}\propto b_+\propto\sin\theta$ between $\ket{\tilde{1}-,\Uparrow^\prime}$ and $\ket{\tilde{2}-,\Downarrow^\prime}$ then results from the magnetic interaction between the majority $\ket{\pm 3/2}$ component of one pseudo-spin with the minority $\ket{\pm 1/2}$ component of the other [Eq. (\ref{eqHz21})]. However, when $b_z\simeq 0$, $\ket{\Uparrow^\prime}$ and $\ket{\Downarrow^\prime}$ become balanced mixtures of the $\ket{\Uparrow}$ and $\ket{\Downarrow}$ states. In these conditions, the Larmor frequency shows a minimum and $\ket{\tilde{2}-,\Downarrow^\prime}$ gets decoupled from $\ket{\tilde{1}-,\Uparrow^\prime}$, as evidenced by the anti-diagonal form of Eq. (\ref{eqHz21}). In other words, both the Zeeman splitting between $\ket{\tilde{1}-}$ states and the coupling between $\ket{\tilde{1}-}$ and $\ket{\tilde{2}-}$ are now driven by the $\propto b_\pm$ interaction between the $\ket{\pm 3/2}$ and $\ket{\pm 1/2}$ envelopes; since the Zeeman-split states defined by Eqs. (\ref{eqprimespin1}) and (\ref{eqprimespin2}) block-diagonalize this interaction (within the $\ket{\Uparrow^\prime}$ and $\ket{\Downarrow^\prime}$ subspaces), $H_{\rm z}$ can not mix $\ket{\Uparrow^\prime}$ and $\ket{\Downarrow^\prime}$ states any more. This gives rise to the dip $G(\theta)$ in the orientational dependence of the Rabi frequency. The dependence of the Rabi frequency on $\varphi$ appearing at higher orders arises from weak confinement anisotropies in the $(xy)$ plane.

\RED{Such anisotropies of the Rabi frequency are ubiquitous in spin-orbit mediated Rabi oscillations, even for electrons.\cite{Stano08,Corna18}} Even if SOC is not explicit in the Luttinger-Kohn Hamiltonian, the present Rabi oscillations result from its action on the $J=3/2$ and $J=1/2$ hole multiplets.\cite{Kloeffel11,Kloeffel13,Kloeffel18} In the absence of SOC, the spin of the holes decouples from their real-space motion so that electrically-driven Rabi oscillations are not possible. The coupling between spin and real space motion in the Luttinger-Kohn Hamiltonian is obvious when writing the total hole wavefunction as a spinor (expanding the physical up and down spin components of the $\ket{j_z}$ Bloch functions).\cite{Luttinger55,KP09} The Luttinger-Kohn Hamiltonian assumes that the splitting $\Delta$ between the $J=3/2$ and $J=1/2$ multiplets is so large that the latter can be dropped out. The physics of low-energy holes then becomes independent on the actual strength of the SOC. The interactions with the nearby $J=1/2$ bands at finite $\Delta$ can be accounted for in the six-bands $\vec{k}\cdot\vec{p}$ model\cite{Dresselhaus55}; However we have checked numerically that they do not make a significant difference in the behavior of the holes for all materials considered in the following.

We would finally like to point out some specificities and limitations of the present model. First, the situation described here is a paradigm of ``$g$-tensor magnetic resonance\cite{Kato03}'' ($g$-TMR) in a strongly anharmonic potential (as discussed in Ref. \onlinecite{Venitucci18}). In this scenario, the Rabi oscillations result from changes in the shape of the qubit wave function driven by the RF electric field (and can be related to the electrical dependence of the principal $g$-factors of the qubit, although we did not follow this approach here). Second, the present model does not account for the orbital correction $\Delta g_z$ on the principal $g$-factor $g_z$ (see Refs. \onlinecite{Watzinger16} and \onlinecite{Venitucci18}) that results from the coupling between $n_z=1$ and $n_z=2$ envelopes by the $S$ term in the Luttinger-Kohn Hamiltonian. This correction has, actually much more impact on the Larmor than on the Rabi frequency in the thin dot limit. 

\subsection{Effects of \RED{quantum dot} orientation and material choice}

In this section, we discuss the impact of the \RED{quantum dot} orientation and material on the speed of the Rabi oscillations. 

As shown by Eq. (\ref{eqfr4}), and discussed in the previous section, the Rabi frequency (at given static electric and magnetic fields) scales primarily with $\zeta_{[110]}$ in the thin dot limit. This parameter does, therefore, adequately characterize the dependence of the Rabi frequency on the choice of box material.

It is also instructive to look at other \RED{quantum dot} orientations -- in particular $x\parallel[100]$, $y\parallel[010]$, $z\parallel[001]$ (the box hence being rotated by $45^\circ$ around the $z$ axis). In that orientation, the Luttinger-Kohn and Zeeman Hamiltonians are the same apart for the $R$ term that becomes:
\begin{equation}
R=\frac{\hbar^2}{2m_0}\sqrt{3}\Big[-\gamma_2\Big(k_x^2-k_y^2\Big)+2i\gamma_3k_xk_y\Big]\,,
\end{equation}
namely $\gamma_2$ and $\gamma_3$ have been interchanged with respect to Eq. (\ref{eqR}). Accordingly, $\gamma_3|\kappa|$ is simply replaced with $\gamma_2|\kappa|$ on the numerator of Eqs. (\ref{eqfr2}) and (\ref{eqfr4}). The Rabi frequency of the box then primarily scales with:
\begin{equation}
\zeta_{[100]}=\frac{\gamma_2|\kappa|}{\gamma_2(\gamma_1+\gamma_2)^2}=\frac{|\kappa|}{(\gamma_1+\gamma_2)^2}\,.
\label{eqzeta100}
\end{equation}
This expression for $\zeta_{[100]}$ outlines the fact that the heavy- and light-hole coupling by lateral confinement is proportional to $\gamma_2$ in this ``$[100]$'' orientation rather than to $\gamma_3$ in the former ``$[110]$'' orientation. This is not expected to make a significant difference in almost isotropic materials such as Ge or III-V 's (where $\gamma_3/\gamma_2\simeq 1$), but is decisive in Si (where $\gamma_3/\gamma_2\simeq 5$). In order to highlight trends among materials, we give $\zeta_{[110]}$ and $\zeta_{[100]}$ for a set of representative materials (Si, Ge and a few III-V's) in Table \ref{tabmat}.

In general, the smaller the bandgap, the larger the Luttinger parameters (smaller hole masses) but the larger $\kappa$. This partly compensates the detrimental effect of the $(\gamma_1+\gamma_2)^2$ factor on the denominators of $\zeta_{[110]}$ and $\zeta_{[100]}$. As a matter of fact, silicon, with its heavier hole masses, but very small $\kappa$ is definitely not the best choice of material for a $[100]$-oriented hole qubit -- although the latter perform, anyway, always worse than $[110]$-oriented hole qubits since $\gamma_3/\gamma_2>1$ for all materials. However, $[110]$-oriented Si qubits, which take advantage of the strong anisotropy of the valence band of Si,\cite{Kloeffel18} show the fastest Rabi oscillations at given static electric and magnetic fields, despite weaker SOC. Indeed, as discussed in section \ref{subsectionPhysics}, the effects of direct SOC\cite{Kloeffel11,Kloeffel13,Kloeffel18} within the heavy- and light-hole manifold become independent on its strength on energy scales much smaller that the spin-orbit splitting $\Delta$. The comparison between materials shall, however, be preferably made at different magnetic fields but at the same Larmor frequency \RED{$f_L\propto|\kappa|$}, which sets the time scale for the intrinsic dynamics of the qubit and the RF circuitry. Also, the comparison is fairer at different $E_0$ but same mixing strength \RED{$\lambda\propto E_0(\gamma_1+\gamma_2)^{-1}$ [Eq. (\ref{eqlambda})]. Indeed, the holes respond stronger to the static electric field when their mass increases, hence reach the same $s$ and $p_y$ envelopes mixing at lower $E_0$. This is also supported by the expression of the optimal $E_{\rm max}^{(0)}\propto(\gamma_1+\gamma_2)$ [Eq. (\ref{eqEmax0})]. We therefore introduce:}
\begin{equation}
\zeta^\prime=\zeta\frac{\gamma_1+\gamma_2}{|\kappa|}\,,
\label{eqzetap}
\end{equation}
namely $\zeta_{[110]}^\prime=(\gamma_3/\gamma_2)\times 1/(\gamma_1+\gamma_2)$ for the $[110]$ orientation and $\zeta_{[100]}^\prime=1/(\gamma_1+\gamma_2)$ for the $[100]$ orientation. With that figure of merit, $[110]$-oriented Si devices remain by far the best choice for a hole spin qubit. We have checked that this conclusion still holds when solving the model in a converged basis set, as well as for more realistic device layouts such as those investigated in Ref. \onlinecite{Venitucci18}. The effects of strains and the case of light-hole qubits are discussed in Appendix \ref{AppendixStrains}.

\section{Conclusions}

To conclude, we have investigated a simple particle-in-a-box model for a hole spin qubit subjected to static electric and magnetic fields and to a radio-frequency electric field that drives Rabi oscillations. We have derived analytical equations for the Rabi frequency in the regime where the Rabi oscillations result from the coupling of the qubit states with a single \RED{excited state}. These equations highlight the dependence of the Rabi frequency on the dimensions and structural orientation of the \RED{quantum dot}, and on the host material parameters. In particular, we show that thin $[110]$-oriented box on $(001)$ substrate perform better than thin $[100]$-oriented box because they can leverage on the anisotropy of the valence band. In this respect, silicon, which displays the most anisotropic valence band among conventional diamond and zinc-blende semiconductors, shows the best opportunities for fast Rabi oscillations in this regime, despite small spin-orbit coupling. The trends outlined by this simple model have been verified in more realistic device layouts close to the silicon-on-insulator devices investigated in Refs. \onlinecite{Venitucci18}.

\begin{acknowledgments}
We thank Alessandro Crippa and Jing Li for a careful reading of the manuscript. This work was supported by the European Union's Horizon 2020 research and innovation program (grant agreement No 688539 MOSQUITO), and by the French National Research Agency (ANR project ``MAQSi'').
\end{acknowledgments}

\appendix

\section{Effects of biaxial strain and Rabi frequency of the light-hole qubit}
\label{AppendixStrains}

We consider an in-plane biaxial strain $\varepsilon_{xx}=\varepsilon_{yy}=\varepsilon_\parallel$, $\varepsilon_{zz}=\varepsilon_\perp=-\nu\varepsilon_\parallel$, where $\nu=2c_{12}/c_{11}$ is the biaxial Poisson ratio and $c_{11}$, $c_{12}$ are the elastic constants of the box material. In the $\{\ket{+\frac{3}{2}},\ket{+\frac{1}{2}},\ket{-\frac{1}{2}},\ket{-\frac{3}{2}}\big\}$ basis set, the Bir-Pikus strain Hamiltonian\cite{Bir74} reads:
\begin{equation}
H_{\rm BP} = 
\begin{pmatrix}
\Delta E_{\rm HH}  & 0 & 0 & 0 \\
0 & \Delta E_{\rm LH} & 0 & 0 \\
0 & 0 & \Delta E_{\rm LH} & 0 \\
0 & 0 & 0 & \Delta E_{\rm HH} \\
\end{pmatrix}\,,
\end{equation}
where:
\begin{subequations}
\begin{align}
\Delta E_{\rm HH}&=\Big[(\nu-2)a_{v}-(\nu+1)b_{v}\Big]\varepsilon_\parallel \\
\Delta E_{\rm LH}&=\Big[(\nu-2)a_{v}+(\nu+1)b_{v}\Big]\varepsilon_\parallel\,,
\end{align}
\end{subequations}
and $a_{v}$, $b_{v}$ are the hydrostatic and uniaxial deformation potentials of the box. Strains therefore rigidly shift pure heavy-hole states with respect to pure light-hole sates. They can be accounted for by replacing $P$ by $P+(\nu-2)a_{v}\varepsilon_\parallel$ and $Q$ by $Q-(\nu+1)b_{v}\varepsilon_\parallel$ in Eq. (\ref{eqHLK}).

In the minimal basis set $\cal{B}$, there exists a relation between the total Hamiltonian $\mathcal{H}_{\rm tot}$ (including electric and magnetic fields) at zero and finite strains. We first notice that $\mathcal{H}_{\rm tot}$ depends on $L_z$ only through the variable $\eta\equiv L_z^{-2}$, \RED{and that}:
\begin{equation}
\mathcal{H}_{\rm tot}(\varepsilon_\parallel, \eta)=\mathcal{H}_{\rm tot}(0, \eta^\prime)+(\nu-2)a_{v}\varepsilon_\parallel\,,
\end{equation}
where:
\begin{equation}
\eta^\prime-\eta=\frac{1}{L_z^{2\prime}}-\frac{1}{L_z^2}=\frac{m_0(\nu+1)b_v}{\hbar^2\pi^2\gamma_2}\varepsilon_\parallel\,.
\label{eqLz2p}
\end{equation}
Therefore, biaxial strain amounts to a change of the squared height of the box:
\begin{equation}
f_R(\varepsilon_\parallel,\eta)=|f_R(0,\eta^\prime)|\,.
\end{equation}
In particular, small compressive (resp. tensile) biaxial strain is equivalent to a decrease (resp. increase) of $L_z^2$ (as $b_v$ is typically negative). Note that $L_z^{2\prime}$ might formally diverge then become negative at large enough tensile strain. Positive and negative $L_z^{2\prime}$ yield the same Rabi frequencies at order $L_z^2/L_x^2$ and $L_z^2/L_y^2$ [Eqs. (\ref{eqfr2}) and (\ref{eqfr2lh}) below], yet not at higher orders. The above scaling relation applies to any pair of qubit states. 

\begin{figure}
\includegraphics[width=0.95\columnwidth]{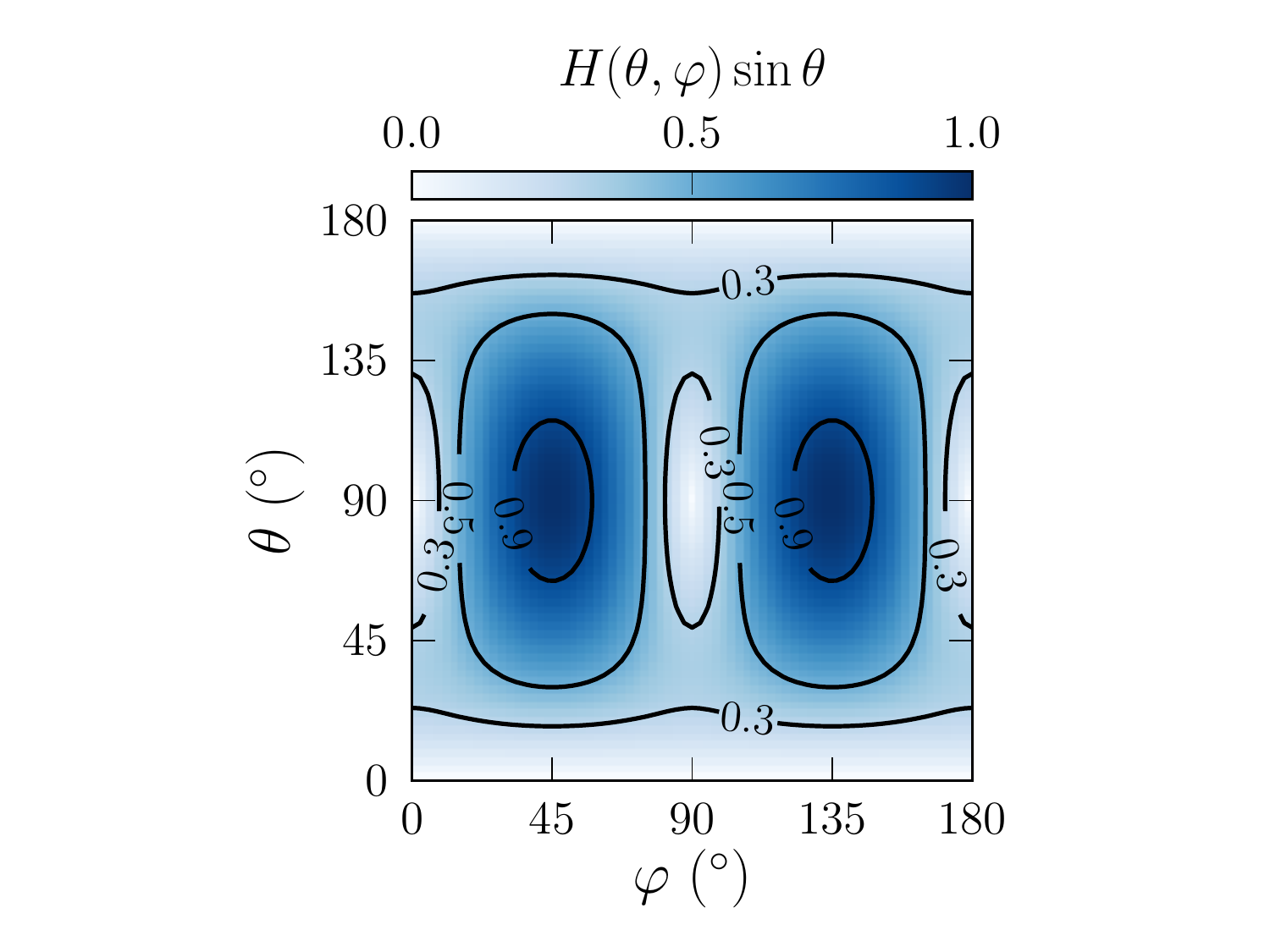} \\
\caption{The function $H(\theta,\varphi)\sin \theta$ characterizing the dependence of the Rabi frequency of the light-hole qubit on the orientation of the magnetic field.}
\label{figLH}
\end{figure}

It must be kept in mind, though, that the ground-state will switch from a mostly heavy- to a mostly light-hole character at large enough tensile strain. In thin dots, the light-hole states at zero strain are the $\ket{1+}$ states. The Rabi frequency of this pair is actually, to order $L_z^2/L_y^2$:
\begin{equation}
f_R^{(2)}=\frac{2^8m_0e^3}{3^4\pi^9\hbar^4}B|E_0|E_{\rm ac}\frac{\gamma_3|\kappa|}{\gamma_2(\gamma_1-\gamma_2)^2}L_y^6\frac{L_z^2}{L_y^2}H(\theta,\varphi)\sin\theta\,,
\label{eqfr2lh}
\end{equation}
with:
\begin{equation}
H(\theta,\varphi)=\sqrt{\frac{1+4\tan^2\theta\sin^2 2\varphi}{1+4\tan^2\theta}}\,.
\end{equation}
The function $H(\theta,\varphi)$ is plotted in Fig. \ref{figLH}. The angular dependence is different from the heavy-hole $\ket{1-}$ pair but the prefactor is the same as Eq. (\ref{eqfr2}) with $(\gamma_1+\gamma_2)^2$ replaced by $(\gamma_1-\gamma_2)^2$ in the denominator. The Rabi frequency may, therefore, be slightly larger for the light-hole than for the heavy-hole states (at same $L_x$, $L_y$, small enough $L_z$ and strains). This results from the fact that ``heavy-holes'' along $z$ (with mass $m_z=m_0/(\gamma_1-2\gamma_2)$) are actually ``light'' in the $(xy)$ plane (with mass $m_{xy}=m_0/(\gamma_1+\gamma_2)$), while ``light-holes'' along $z$ ($m_z=m_0/(\gamma_1+2\gamma_2)$) are ``heavy'' in the $(xy)$ plane ($m_{xy}=m_0/(\gamma_1-\gamma_2)$), hence respond stronger to the electric and magnetic fields [see the expressions of $P$ and $Q$ in Eq. (\ref{eqHLK})]. The Rabi frequency of the light-hole pair is maximum for $\theta=90^\circ$, $\varphi=45^\circ$ (modulo $90^\circ$), while the Rabi frequency of the heavy-hole pair is maximum for $\varphi=0^\circ$ (modulo $180^\circ$), but for a polar angle $\theta$ that depends on the dimensions of the qubit. The Larmor frequency is also significantly less anisotropic for the light-hole pair (as $|g_x|\simeq|g_y|\simeq 4|\kappa|$, $|g_z|\simeq 2|\kappa|$). 

\begin{figure}
\includegraphics[width=0.95\columnwidth]{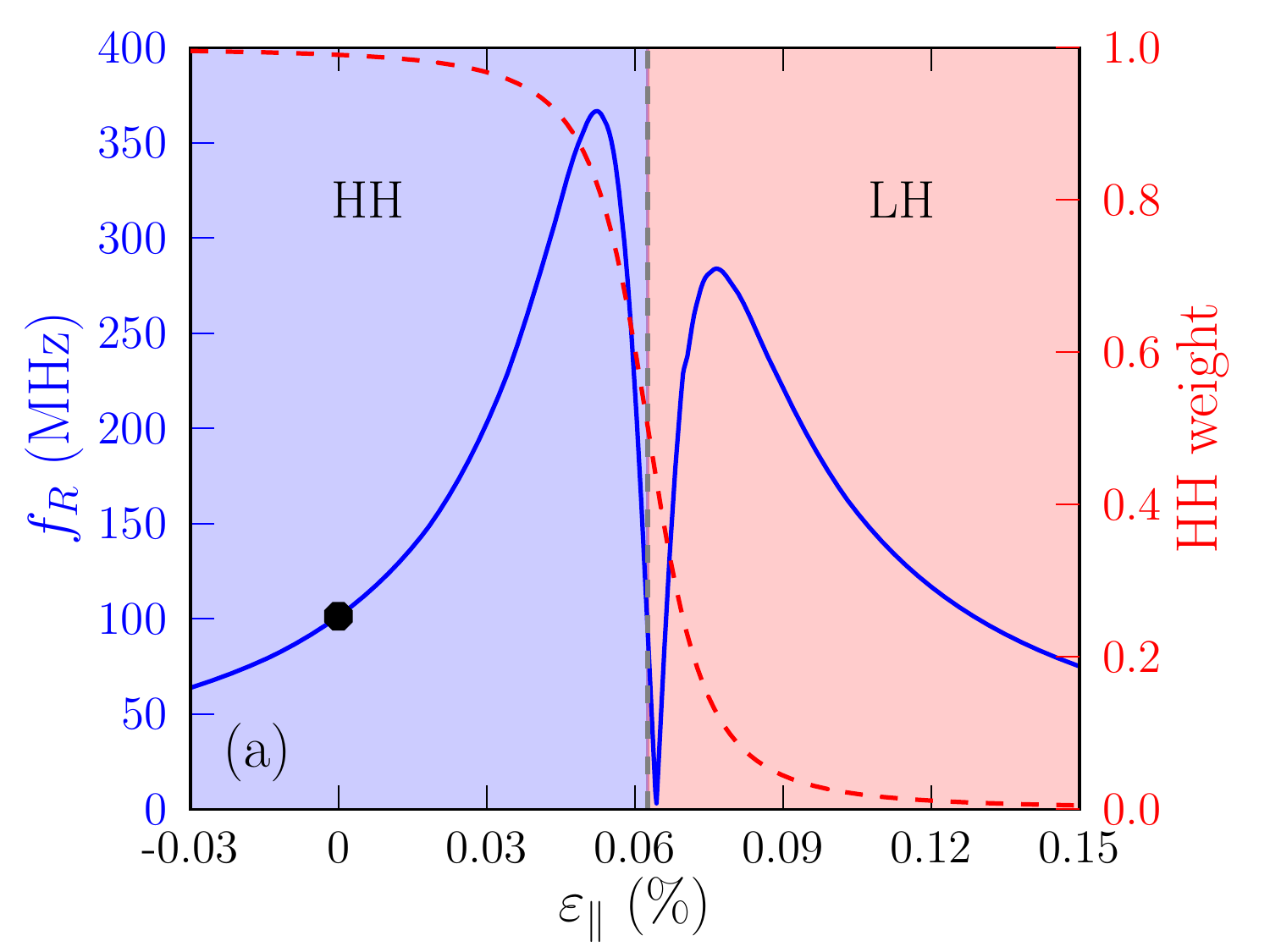} \\
\includegraphics[width=0.95\columnwidth]{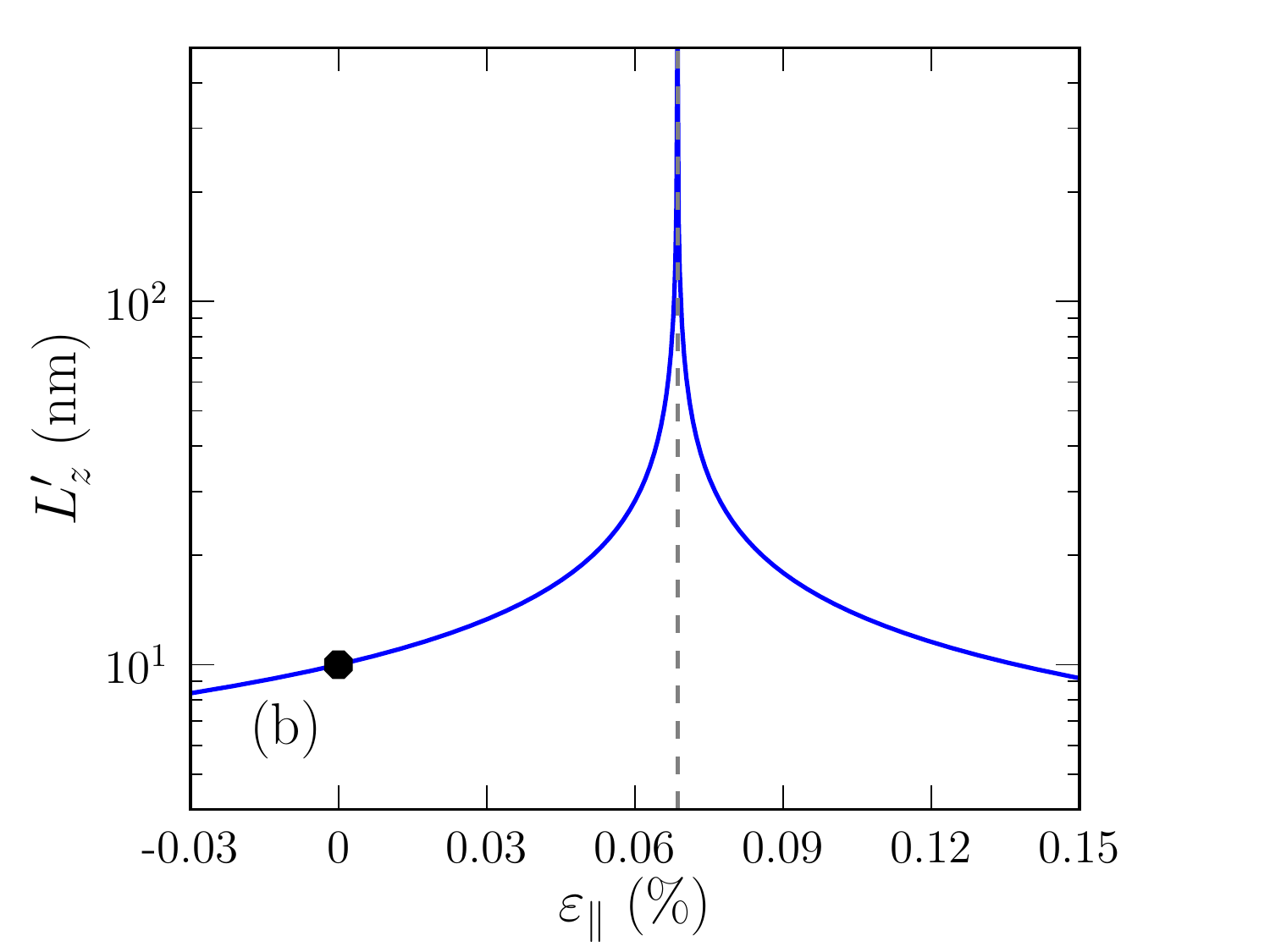} 
\caption{(a) Heavy-hole composition (dashed red line, right scale) and Rabi frequency (solid blue line, left scale) of the ground-state pair as a function of $\varepsilon_\parallel$ in a silicon box with sides $L_x=40$ nm, $L_y=30$ nm and $L_z=10$ nm. It is computed from the exact solution of the Hamiltonian in the basis set ${\cal B}$. The RF electric field is $E_{\rm ac}=0.03$ mV/nm, and the magnetic field $B=1$ T is oriented along the optimal direction for each $\varepsilon_\parallel$. The transition from a mostly heavy-hole (HH) to a mostly light-hole (LH) ground-state takes place at $\varepsilon_\parallel=\varepsilon_\parallel^*=0.0625$ \%. (b) Effective $L_z^\prime=\sqrt{|L_z^{2\prime}|}$ as a function of $\varepsilon_\parallel$ [Eq. (\ref{eqLz2p})]. $L_z^{2\prime}$ diverges at $\varepsilon_\parallel=\varepsilon_\parallel^\infty=0.0686$ \%; it is positive for $\varepsilon_\parallel<\varepsilon_\parallel^\infty$, and negative for $\varepsilon_\parallel>\varepsilon_\parallel^\infty$. The reference point $\varepsilon_\parallel=0$ is highlighted by a black dot on both plots.}
\label{figepsilon}
\end{figure}

The heavy-hole composition $h_1^2$ and the Rabi frequency of the ground-state pair are plotted as a function of $\varepsilon_\parallel$ in Fig. \ref{figepsilon}a, in a silicon box with sides $L_x=40$ nm, $L_y=30$ nm and $L_z=10$ nm. They are computed from the exact solution of the Hamiltonian in the basis set ${\cal B}$ ($\nu=0.77$, $b_v=-2.1$ eV). The RF electric field is $E_{\rm ac}=0.03$ mV/nm, and the magnetic field $B=1$ T is oriented along the optimal direction (maximum $f_R$) for each $\varepsilon_\parallel$. For $\varepsilon_\parallel<\varepsilon_\parallel^*=0.0625$ \%, the qubit states have a mostly heavy-hole character, while for $\varepsilon_\parallel>\varepsilon_\parallel^*$, they have a mostly light-hole character. The Rabi frequency decreases at large compressive or tensile strain because the heavy- and light-hole components get strongly split, which suppresses the necessary heavy- and light-hole mixings in the qubit and excited states (equivalently, $L_z^{2\prime}\to0$, as shown in Fig. \ref{figepsilon}b). The Rabi frequency also exhibits a peak split by a dip near (but not exactly at) the transition strain $\varepsilon_\parallel=\varepsilon_\parallel^*$. This peak results from an increase of the effective $L_z^{2\prime}$ (stronger heavy- and light-hole mixing), although neither Eq. (\ref{eqfr2}) nor Eq. (\ref{eqfr2lh}) are actually applicable in this range. The dip is centered at the strain $\varepsilon_\parallel=\varepsilon_\parallel^0=0.0643$ \% where $h_1=h_2$, $l_1=l_2$.\footnote{The dip is centered either at $h_1=h_2$, $l_1=l_2$ or at $h_2=-l_1$, $h_1=l_2$ depending on the dimensions of the box.} The eigenstates of $H_{\rm LK}$ can then all be factored as the products of single envelopes by mixed heavy- and light-hole Bloch functions. Since either the envelope or the Bloch function of the different states must be orthogonal, the qubit and excited states can not be coupled by both $E_{\rm ac}$ and $H_{\rm z}$ any more in Eq. (\ref{eqfrabisos}). The dip is partly smoothed out (but does not disappear) in larger basis sets. Therefore, hole spin qubits turn out to be very sensitive to strains, and the range of $\varepsilon_\parallel$ that really enhance the Rabi frequency is pretty narrow. Overall, the Rabi frequency remains larger for the mostly heavy-hole than for the mostly light-hole qubit near the peak [where, again, Eqs. (\ref{eqfr2}) and (\ref{eqfr2lh}), which suggest the opposite behavior, do not hold].

\section{Equations for $\Pi_{\tilde{1}+}$, $\Pi_{\tilde{2}+}$ and $\Pi_{\tilde{2}-}$}
\label{AppendixSOS}

The equations for $\Pi_{\tilde{2}-}$ are:
\begin{align}
BE_0\Pi_{\tilde{2}-}=&\frac{D_1}{E_{1-}-E_{2-}}\times \nonumber \\
\Big\{\lambda_{2-}^{1-}\Big[&-4\alpha\beta(Z_1^{(2)}-Z_1^{(1)}) \nonumber \\
&-2\beta^2(Z_2^{(2)}-Z_2^{(1)}) \nonumber \\
&+2\alpha^2(Z_2^{(2)*}-Z_2^{(1)*})\Big] \nonumber \\
+\lambda_{2+}^{1-}\Big[&-4\alpha\beta Z_3^{(2)}-2\beta^2Z_4^{(2)}+2\alpha^2Z_4^{(2)*}\Big] \nonumber \\
+\lambda_{1+}^{2-}\Big[&-4\alpha\beta Z_3^{(1)}-2\beta^2Z_4^{(1)}+2\alpha^2 Z_4^{(1)*}\Big]\Big\}\,,
\end{align}
with:
\begin{equation}
D_1=\braket{2-,\Uparrow|y|1-,\Uparrow}=-\frac{16L_y}{9\pi^2}(h_1h_2+l_1l_2)
\end{equation}
and:
\begin{subequations}
\begin{align}
Z_1^{(i)}&=\braket{i-,\Uparrow|H_{\rm z}|i-,\Uparrow} \nonumber \\ 
&=\kappa\mu_BB(3h_i^2-l_i^2)b_z \\
Z_2^{(i)}&=\braket{i-,\Uparrow|H_{\rm z}|i-,\Downarrow} \nonumber \\ 
&=2\kappa\mu_BB(\sqrt{3}h_il_i b_-+l_i^2 b_+)\\
Z_3^{(i)}&=\braket{i-,\Uparrow|H_{\rm z}|i+,\Uparrow} \nonumber \\ 
&=-4\kappa\mu_BBh_il_ib_z\\
Z_4^{(i)}&=\braket{i-,\Uparrow|H_{\rm z}|i+,\Downarrow} \nonumber \\ 
&=2\kappa\mu_BB\Big[\frac{\sqrt{3}}{2}(h_i^2-l_i^2)b_-+l_i h_ib_+\Big]\,.
\end{align}
\end{subequations}

The equations for $\Pi_{\tilde{2}+}$ are likewise:
\begin{align}
BE_0\Pi_{\tilde{2}+}=&\frac{D_2}{E_{1-}-E_{2+}}\times \nonumber \\
\Big\{\lambda_{2-}^{1-}\Big[&-4\alpha\beta Z_3^{(2)}-2\beta^2Z_4^{(2)}+2\alpha^2 Z_4^{(2)*}\Big] \nonumber \\
+\lambda_{1+}^{2+}\Big[&-4\alpha\beta Z_3^{(1)}-2\beta^2Z_4^{(1)}+2\alpha^2Z_4^{(1)*} \Big] \nonumber \\
+\lambda_{2+}^{1-}\Big[&-4\alpha\beta(Z_5^{(1)}-Z_1^{(1)}) \nonumber \\
&-2\beta ^2(Z_6^{(2)}-Z_2^{(1)}) \nonumber \\
&+2\alpha^2(Z_6^{(2)*}-Z_2^{(1)*})\Big]\Big\}\,,
\end{align}
where:
\begin{equation}
D_2=\braket{2+,\Uparrow|y|1-,\Uparrow}=\frac{16L_y}{9\pi^2}(h_2l_1-h_1l_2) \\
\end{equation}
and:
\begin{subequations}
\begin{align}
Z_5^{(i)}&=\braket{i+,\Uparrow|H_{\rm z}|i +,\Uparrow} \nonumber \\
&=\kappa\mu_BB(3l_i^2+h_i^2)b_z \\
Z_6^{(i)}&= \braket{i+,\Uparrow|H_{\rm z}|i+,\Downarrow} \nonumber \\
&=2\kappa\mu_BB(-\sqrt{3}h_il_i b_-+h_i^2 b_+)\,.
\end{align}
\end{subequations}

Finally, the equation for $\Pi_{\tilde{1}+}$ is:
\begin{align}
BE_0\Pi_{\tilde{1}+}&=\frac{-4\alpha\beta Z_3^{(1)}-2\beta^2Z_4^{(1)}+2\alpha^2 Z_4^{(1)*}}{E_{1-}-E_{1+}} \nonumber \\
&\times\Big[D_1\big(\lambda_{2-}^{1+}+\lambda_{2+}^{1-}\big)+D_2\big(\lambda_{2+}^{1+}-\lambda_{2-}^{1-}\big)\Big]\,.
\end{align}

\end{document}